\documentclass[%
 reprint,
superscriptaddress,
 amsmath,amssymb,
 aps,prmaterials,
floatfix,
]{revtex4-1}

\usepackage{graphicx}
\usepackage{subcaption}
\usepackage{verbatim}
\usepackage{float}
\usepackage{dcolumn}
\usepackage{bm}
\usepackage{epsfig}
\usepackage{amsmath}
\usepackage[colorlinks = true, linkcolor = blue, urlcolor  = blue, citecolor = blue, anchorcolor = blue]{hyperref}

\newcommand{\NEW}[1]{\textnormal{#1}}

\begin{document}

\title{Complex magnetic textures in Ni/Ir$_{n}$/Pt(111) ultrathin films}

\author{P. C. Carvalho}
\author{I. P. Miranda}
\affiliation{Universidade de S\~ao Paulo, Instituto de F\'isica, Rua do Mat\~ao, 1371, 05508-090 S\~ao Paulo, S\~ao Paulo, Brazil}

\author{A. B. Klautau}
\email{aklautau@ufpa.br}
\affiliation{Faculdade de F\'isica, Universidade Federal do Par\'a, CEP 66075-110, Bel\'em, PA, Brazil}

\author{A. Bergman}
\affiliation{Department of Physics and Astronomy, Uppsala University, 75120 Box 516 Sweden}

\author{H. M. Petrilli}
 \affiliation{Universidade de S\~ao Paulo, Instituto de F\'isica, Rua do Mat\~ao, 1371, 05508-090 S\~ao Paulo, S\~ao Paulo, Brazil}

\date{\today}

\begin{abstract}

A combined approach using first-principles calculations and spin dynamics simulations is applied to study Ni/Ir$_{n}$/Pt(111) ($n=0,1,2$) films. The lowest-energy states are predicted to be spin-spirals but with a minute (of the order of a few $\mu$eV/atom) energy difference with skyrmionic states. The spontaneous low temperature  skyrmions, with $\sim15$ nm to $\sim35$ nm size, arise from a large Dzyaloshinskii{–}Moriya (DM) and Heisenberg exchange interactions ratio and, in particular, from a large in-plane DM vector component for nearest neighbors. The skyrmions become larger and more dispersed with the enhancement of the Ir buffer thickness.
Also, with increasing $n$, the skyrmions stability decrease when an external magnetic field is applied or  the temperature is raised. For $n=0$ and $n=1$, we found that metastable skyrmioniums can occur, which are characterized by a slightly lower stability with respect to the external fields
and larger critical currents, compared to skyrmions.
\end{abstract}

\maketitle

\section{\label{sec:introduction}INTRODUCTION}

Magnetic skyrmions are topologically protected chiral spins structures \cite{Fert2013,Jiang2017, Bogdanov1989, Bogdanov1999} and, although the first experimental observation in MnSi bulk \cite{Neubauer2009,Pappas2009,Muhlbauer915} is relatively recent, a vast research has already been made exploring the existence of these nanostructures in different materials. 
The search for faster magnetic storage devices and a larger storage capacity, among other features, has drawn attention to new magnetic nanostructures \cite{Fert2013,Zhang2015,Gobel2019}.
For instance, the presence of skyrmions has been reported in several metallic multilayers \cite{Romming2013,Herve2018,Legrand2019,Moreau-Luchaire2016,Boulle2016,Soumyanarayanan2017}.
Other  magnetic structures that stand out in recent literature are magnetic skyrmioniums (or $2\pi$-skyrmions) \cite{Finazzi2013}. These are doughnut-like nontopological solitons, which have been proposed and studied theoretically \cite{Komineas2015,Vigo2020}, as well as experimentally verified \cite{Zhang2018,Finazzi2013}. They have been attracting much attention due to  its  $q=0$ \cite{Zhang2016} topological charge. This characteristic makes these structures unaffected by the so-called skyrmion Hall effect (SkHE) \cite{Zang2011, Jiang20172,Litzius2017,Litzius2020}, which, if present, forces the skyrmion to deviate from the current direction and may represent a challenge for future design of magnetic devices \cite{Purnama2015}, specially for racetrack applications, as it can cause an annihilation of the skyrmion at the edges.

The exact conditions for the formation of skyrmions and skyrmioniums are not yet completely known where, among other interactions  \cite{Souvik2020}, long range dipole-dipole \cite{Montoya2017,Bernand2020}, anisotropy \cite{Wilson2014}, frustrated exchange \cite{Malottki2017,Meyer2019}, the Dzyaloshinskii{–}Moriya \cite{Dzyaloshinsky1958,Moriya1960} (DM), as well as their interconnections,  play a role. 
The DM is an anti-symmetric interaction which is non-zero in the presence of broken inversion symmetry. Its strength can be related to a high spin orbit coupling (SOC) \cite{Fert2013} and it has been shown that it can also be associated with the degree of $3d$-$5d$ hybridization around the Fermi energy for certain materials \cite{Belabbes2016}. The formation of skyrmions has been explored in the literature regarding the DM interaction strength.
For instance, Moreau-Luchaire et al. \cite{Moreau-Luchaire2016} experimentally observed room temperature skyrmions in Ir/Co/Pt asymmetric multilayers, stabilized by a large DM interaction.

The DM interaction can be the main mechanism in the formation of skyrmions, in particular, in the absence of frustrated exchange interactions and for skyrmions smaller than $\approx$100 nm, where the dipole-dipole interactions are negligible \cite{Nagaosa2013, Fert2017}. While the DM induces a chirality sense to the system, the Heisenberg exchange \cite{Liechtenstein1987} interactions can favor a collinear (or non-collinear, through magnetic frustrations) alignment of spins, so the formation of skyrmions in magnetic films can be understood in terms of a competition between these interactions \cite{Fert2017,Herve2018}. Therefore, interfaces composed by magnetic materials and high SOC elements are good candidates to search for skyrmions.

Among the systems composed by high SOC elements and magnetic materials, the study of Ni/Pt(111) and Ni/Ir(111) ultrathin films \cite{Bornemann2012,Belabbes2016} stands out, regarding its electronic structure, exchange and DM interactions.
It has been shown that the strength of the DM interaction, for nearest-neighbors, is smaller in the case of ultrathin Ni/Pt(111) layers than in systems with other $3d$ transition metals such as Fe/Pt(111) and Co/Pt(111). However, it was also revealed that, since the exchange and DM interactions were often of the same order of magnitude, the emergence of complex noncollinear configurations in the ground state of Ni clusters on Pt(111) and Ir(111) surfaces \cite{Bornemann2012} is feasible. 
Besides that, the addition of spacer layers between the magnetic and non-magnetic elements can lead to different magnetic behaviors as shown in Ref. \cite{Vida2016}, for Co/Ir$_{n}$/Pt(111) ($0\leq n\leq6$) ultrathin films.
To the best of our knowledge the magnetic properties  of Ni/Ir$_{n}$/Pt(111) ultrathin films nor the importance of the DM direction into the stability of these complex noncollinear configurations have not yet been reported.

Here, we use atomistic spin dynamics simulations in combination with first principles calculations to predict the magnetic configurations of Ni/Ir$_{n}$/Pt(111) ultrathin films, with \textit{n} = 0, 1 and 2. By using density functional theory (DFT), we calculate Heisenberg exchange and strength of the DM interactions as well as the direction of the DM vectors. These quantities are used as input in atomistic 
spin dynamics simulations aiming to predict the Ni layers magnetic ground state. Skyrmions stabilities, regarding temperature and magnetic fields, are investigated through the Ni/Ir$_{n}$/Pt(111) phase diagram. Furthermore,  skyrmions and skyrmioniums shape, as well as their velocities under applied currents, are also analyzed.

\section{\label{sec:comp_methods}COMPUTATIONAL METHODS}

Electronic structure calculations were performed with the real-space linear-muffin-tin-orbital within the atomic sphere approximation (RS-LMTO-ASA) method \cite{Pessoa1992,Peduto1991,Klautau1999,Rodrigues2016,BezerraNeto2013,Cardias2016,Kvashnin2016} to obtain the magnetic ground state properties. The RS-LMTO-ASA has been proven to be suitable to describe complex magnetic systems \cite{Kvashnin2016,Cardias2016,Rodrigues2016,BezerraNeto2013,klautauMagneticPropertiesCo2004,bergmanNoncollinearMagnetisationClusters2006,frota-pessoaInfluenceInterfaceMixing2002}.
To obtain the electronic density, the Haydock recursion \cite{Haydock1980} method with the Beer-Pettifor terminator \cite{Beer1984} was used.  Fully relativistic first principles calculations were performed with the local spin density approximation (LSDA) \cite{Barth1972}, for the exchange-correlation functional, and the recursion cut-off set to $LL=22$. The exchange coupling ($J_{ij}$) and DM vector ($\vec{D}_{ij}$), between atomic sites \textit{i} and \textit{j} located at the surface, were calculated for a ferromagnetic configuration using the RS-LMTO-ASA code \cite{Sonia2000,Liechtenstein1987,Cardias2020}. Since the $J_{ij}$ and $|\vec{D}_{ij}|$ values can be different for $(i,j)$ pairs of atoms in the same neighboring shell, due to the influence of the lower layer displacement that follows the fcc(111) stacking,  the average was considered, for each interatomic distance.

We studied Ni/Ir$_{n}$/Pt(111), where 0 $\leq$ \textit{n} $\leq$ 2 
is the number of Ir spacer layers. A cluster of  $\sim 6,000$ atoms was built to simulate  the fcc(111) ultrathin layer in real space with the experimental Pt(111) lattice parameter 3.92 \AA. The system is composed by: one layer of atoms  represented by empty spheres to mimic vacuum; one layer of Ni atoms (denoted as \textit{s} layer) below the empty spheres; \textit{n} layers of Ir atoms adjacent to the Ni layer, with 0 $\leq$ \textit{n} $\leq$ 2 (denoted as \textit{s-1}, \textit{s-2} and  \textit{s-3} layers); five layers of Pt atoms below the Ni layer (when \textit{n} = 0) or the Ir layers (when \textit{n} $\neq$ 0). 

The atomistic spin dynamics \cite{Antropov1996} (ASD) simulations were performed with the UppASD code \cite{Eriksson2017,Skubic2008}. Firstly, the system is brought to its ground state through Monte Carlos (MC) simulations, from a temperature $T$ larger than the Curie temperature $T_{\textnormal{C}}$ ($T>T_{\textnormal{C}}$) to $T\sim0$ K. In this initial phase, the atomic spin directions are varied in order to obtain the spin configuration with minimum energy. With the ground state configuration, the Landau-Lifshitz-Gilbert (LLG) equation \cite{Gilbert1955,Eriksson2017,Slonczewski1996} is used to obtain the time evolution of each atomic magnetic moment $i$ ($\vec{m}_{i}$) and relax the system, accounting for the spin-orbit torque (SOT) mechanism \cite{Slonczewski1996,Manchon2019}:

\begin{equation}\label{eq:llg}
\begin{split}
\frac{d\vec{m}_{i}}{dt}=-\frac{\gamma}{1+\alpha^{2}}\vec{m}_{i}\times\left[\vec{B}^{i}_{\textnormal{eff}}+\frac{\alpha}{m_{i}}(\vec{m}_{i}\times\vec{B}^{i}_{\textnormal{eff}})\right]\\+{\tau_{DL}}\left[(\vec{m}_{i}\times\vec{s})\times\vec{m}_{i}\right]-{\tau_{FL}}(\vec{m}_{i}\times\vec{s}),
\end{split}
\end{equation}

\noindent where $\vec{B}^{i}_{\textnormal{eff}}=-\frac{\partial\mathcal{H}}{\partial\vec{m}_{i}}+\vec{b}_{i}(T)$ is the effective field acting over the $i$th site, $\vec{b}_{i}(T)$ is a stochastic field to consider temperature $T$ effects by using Langevin dynamics \cite{Skubic2008}, $\alpha$ is the Gilbert damping parameter, $\gamma$ is the gyromagnetic ratio, $\vec{s}$ is the polarization direction of the spin current, and $\tau_{DL}$ ($\tau_{FL}$) are the damping-like (field-like) coefficient. Here, $\tau_{FL}$ is set to zero due to the negligible influence on the dynamics \cite{Tomasello2014,Manchon2019,Zhang2015}. In turn, the damping-like parameter, $\tau_{DL}$, is defined as $\tau_{DL}=\frac{g\mu_{B}\theta_{\textnormal{SH}}}{2eM_{s}t_{F}}$, being $g$ the Landé $g$-factor, $e$ the electron charge, $\theta_{\textnormal{SH}}$ the  spin Hall angle, $t_{F}$ the ferromagnetic thickness, and $M_{s}$ the saturation magnetization of the sample (calculated from the obtained magnetic moments). In all calculations, we considered $t_{F}\sim2.26\,$\AA$\,$ and the spin Hall angle of Pt, namely $\theta_{\textnormal{SH}}\sim0.03$ \cite{Tao2018} (assuming that Ir and Ir/Pt might present $\theta_{\textnormal{SH}}$ almost in the same order of magnitude \cite{Saito2021,Fache2020}).

The connection between RS-LMTO-ASA and ASD simulations is made by the Hamiltonian $\mathcal{H}$

\begin{equation}\label{hamil}
\begin{split}
    \mathcal{H} = -\sum_{i \ne j}J_{ij}(\hat{e}_{i}\cdot\hat{e}_{j}) - \sum_{i \ne j}\vec{D}_{ij}\cdot(\hat{e}_{i}\times\hat{e}_{j}) \\
    - \sum_{i}\vec{{B}}_{\textnormal{ext}}\cdot\hat{e}_{i},
\end{split}
\end{equation}

\noindent in which the Heisenberg exchange interaction ($J_{ij}$),  Dzyaloshinskii{–}Moriya interaction ($\vec{D}_{ij}$) and the local magnetic moment  ($\vec{{m}}_{i}={m} \hat{e}_{i}$) are obtained from first principles calculations. Here, we consider the interactions $J_{ij}$ and $\vec{D}_{ij}$ between the first five shells of $3d-3d$ neighbors.
The dipole-dipole interaction and the magnetic anisotropy are neglected \cite{Fert2017,Herve2018} 
\NEW{(see Sec. \ref{subsec:spin_dyn})}.
The external magnetic field $\vec{{B}}_{\textnormal{ext}}$ can be included in the Hamiltonian optionally. The simulated system size is a very sensitive parameter when studying spin-spirals and skyrmions. There can be an energy penalty associated with periodic simulation cells if they are incommensurate with the magnetic textures which can lead to misleading results if not analyzed properly. For these systems, we found that a square lattice with $280\times280$ atomic spins is suitable for simulating the Ni monolayer 
\NEW{(see Sec. \ref{subsec:finite-effects}} 
for more details) and used the calculated damping values from first principles \cite{Miranda2021} of $\alpha=0.070$ (Ni/Pt(111)), $\alpha=0.086$ (Ni/Ir$_1$/Pt(111)), and $\alpha=0.062$ (Ni/Ir$_2$/Pt(111)), which are in agreement with suitable parameters for similar surfaces \cite{Barati2014,Mizukami2011,Bergqvist2013}. We notice that only the Ni layer is simulated, since  the influence of Ir and Pt atoms are included in the computation of the $J_{ij}$ and $|\vec{D}_{ij}|$  interactions between Ni atoms.  The skyrmions stabilities are analysed  through the topological charges $q$ (of each skyrmion) or $Q=\sum_{i}q_{i}$ (of the entire lattice). The topological charge was calculated by analyzing the spherical angles between neighbouring spins \cite{Berg1981}.

\section{\label{sec:results}RESULTS AND DISCUSSION}

\subsection{\label{subsec:abinitio}\textit{Ab initio} calculations}

Table \ref{tab:mommag} presents the local spin magnetic moment ($m$) of a representative atom at each layer in the Ni/Ir$_{n}$/Pt(111) systems studied here. The effect of the insertion of Ir spacer layers is to decrease  the Ni local magnetic moment ($m_{\textnormal{Ni}}$), up to $\Delta m_{\textnormal{Ni}}$ = 0.18 $\mu_{B}$. A similar reduction in $m_{\textnormal{Ni}}$ can also be found  from Ni/Pt(111) to Ni/Ir(111), in the literature \cite{Bornemann2012}. Furthermore, the induced magnetic moment at the  \textit{s-1} layers  in  Ni/Pt(111) and Ni/Ir$_{1}$/Pt(111) are equal, but are smaller in the Ni/Ir$_{2}$/Pt(111) case. The induced moments at the \textit{s-1} layers can be explained by the hybridization of $3d$-$5d$ states (see Appendix \ref{sec:appena:Ni_es}).

\begin{table}[h]
	\centering
	\caption{\small{Local magnetic moment ${m}$ (in $\mu_{B}$/atom) of a typical atom in the first three layers of the Ni/Ir$_{n}$/Pt(111) systems.  The layers are denoted by \textit{s}, \textit{s-1} and \textit{s-2} (see text).}}
\begin{ruledtabular}
	\begin{tabular}{c|cc|cc|cc}
	 &\multicolumn{2}{c|}{$n=0$} & \multicolumn{2}{c|}{$n=1$} & \multicolumn{2}{c}{$n=2$} \\ \hline
	\textit{s} &	Ni & 0.60 & Ni & 0.52 & Ni & 0.42 \\
	\textit{s-1} & Pt & 0.12 & Ir & 0.12  & Ir & 0.06  \\
	\textit{s-2} & Pt & 0.03 & Pt & 0.01  & Ir & -0.01  \\ 
	\end{tabular}
	\label{tab:mommag}
\end{ruledtabular}
\end{table}

\begin{figure}[h]
\centering
\includegraphics[width=\linewidth, height=11cm]{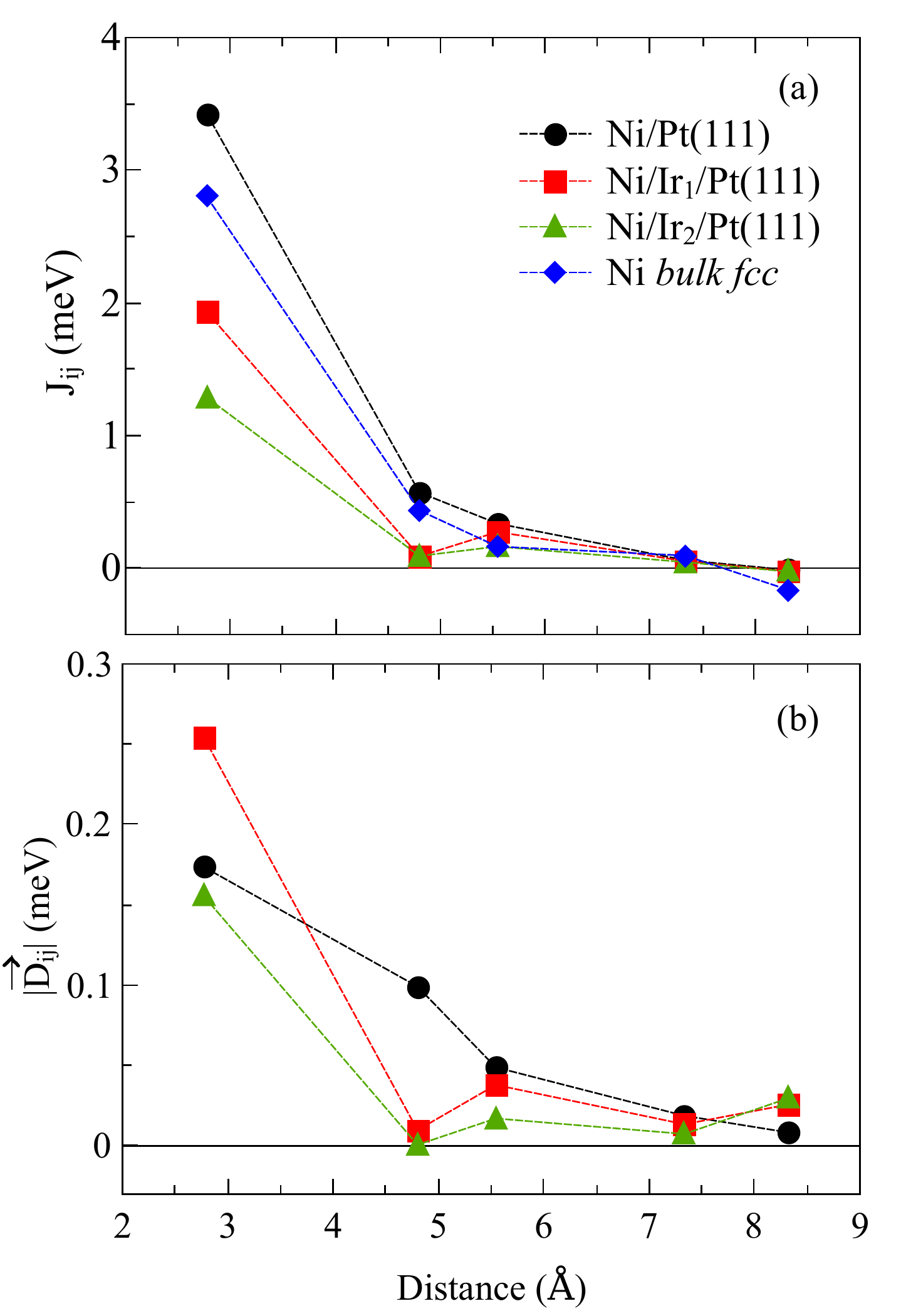}
\caption{\label{fig:jDNi} \small{Ni-Ni (a) Exchange coupling ($J_{ij}$) and (b) DM interaction strength ($|\vec{D}_{ij}|$) as a function of the interatomic distance for Ni/Ir$_{n}$/Pt(111) (\textit{n} = 0, 1 and 2). Data for Ni \textit{bulk fcc} obtained from Ref. \cite{Sonia2000}. The dotted lines are guides to the eyes. Convention symbols are the same for (a) and (b).}}
\end{figure}

 Heisenberg ($J_{ij}$) and  strength of the DM ($|\vec{D}_{ij}|$) interactions between Ni atoms
  in the Ni/Ir$_{n}$/Pt(111) systems
 are shown in Fig.~\ref{fig:jDNi} as a function of the interatomic distance. 
 Ni \textit{bulk fcc} $J_{ij}$'s,  calculated with the same method \cite{Sonia2000}, are also presented.
It can be seen that the successive addition of Ir spacer layers  causes a significant decrease of the Ni-Ni nearest neighbors (NN) exchange coupling ($J_{1}$), following the same $m_{\textnormal{Ni}}$ trend: the largest  $J_{1}$ is for Ni/Pt(111)  (\textit{n} = 0), by adding one Ir spacer layer (\textit{n} = 1) it decreases almost 50$\%$ and for two Ir spacer layers (\textit{n} = 2) the interaction is even smaller. 

This trend (reduction of $J_{1}$ as the number of Ir spacer layers is increased) is in accordance with the comparative behavior  of the extreme cases, Ni/Pt(111) and Ni/Ir(111), reported in the literature \cite{Bornemann2012}. It can also be seen from Fig.~\ref{fig:jDNi}(a) that  $J_{ij}$  decrease significantly for distances larger than 5 \AA. Interestingly,  $J_{1}$ is larger in  Ni \textit{bulk} than in  the ultrathin layers with Ir spacers, but smaller than in Ni/Pt(111).

From Fig.~\ref{fig:jDNi}(b) we observe that, analogous to the exchange coupling, the DM interaction strengh is larger for NN ($D_{1}$) and decreases for further neighbors, with a certain oscillation in the values. Remarkably is the NN interaction for \textit{n} = 1 ($D_{1}$ = 0.25 meV) compared to \textit{n} = 0 and \textit{n} = 2. In order to understand this behaviour we note that three aspects have been suggested to contribute to the DM interaction: 
(\textit{a}) the strength of SOC in a $5d$ substrate;
(\textit{b}) the inversion symmetry breaking; and 
(\textit{c}) the spin polarization between $3d$-$5d$ interface atoms \cite{Belabbes2016}.   The presence of a larger DM interaction for $n=1$, compared with $n=0$, can then be explained by a higher symmetry breaking (more heterogeneous system), although Ir SOC is smaller than Pt, and the induced magnetic moments at the interfaces (\textit{s-1} layer) have the same value (see Table \ref{tab:mommag}).
The induced moment in the $n=2$ case is smaller than for $n=0$ and $n=1$, the SOC is also smaller, and this inclusion of two Ir layers tends to the behaviour of the system in the extreme case of Ni/Ir(111).
Also, from Fig.~\ref{fig:jDNi}(b), we notice that,   Ni/Ir$_1$/Pt(111) has the largest magnitude of DM interaction for NN, and for next-nearest-neighbours (NNN) the DM interaction is much stronger in Ni/Pt(111) compared to the other two systems. Therefore, only $D_{1}$ is relevant in the  $n=1$ and $n=2$ cases. 
The role of these NNN DM interactions in the spin dynamics results will be discussed in \NEW{Sec.} \ref{subsec:phase_diag}.

\begin{figure}[h]
\centering
\includegraphics[width=\linewidth]{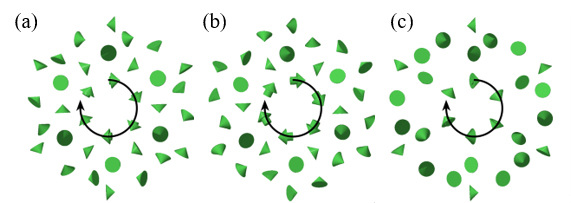}
\caption{\label{fig:dvec_Ni} \small{$\vec{D}_{ij}$ between the central Ni and the Ni atom in the site of the arrow in: (a) Ni/Pt(111), (b) Ni/Ir$_{1}$/Pt(111) and (c) Ni/Ir$_{2}$/Pt(111). Arrow sizes are proportional to the vectors strengths. In (c), missing vectors are due to vanishing interaction strengths.}}
\end{figure}

Besides the strength of the DM vector, the direction is also an important aspect to be analyzed, since for a \textit{fcc}(111) surface the DM direction is not completely determined from the Moriya's symmetry rules \cite{Moriya1960,Crepieux1998}. The image of the hexagonal island with the DM vectors representing the interaction between Ni atoms, here obtained, are shown in Fig.~\ref{fig:dvec_Ni}. We verified that the DM vectors follow the rule $\vec{D}_{ij}\cdot\vec{{R}}$ = 0 \cite{Moriya1960,Crepieux1998}, where $\vec{{R}}$ is the vector connecting the  \textit{i} and \textit{j} sites. The rotational senses of the vectors are represented with  black arrows. We see that adding Ir layers in Ni/Ir$_{n}$/Pt(111) does not change the rotational sense of the DM vectors. Moreover, the vectors are mostly in the plane of the surface, i.e. they have a small out-of-plane component. In Fig.~\ref{fig:dvec_Ni}(c), some arrows are not shown due to a very small DM strength.

\subsection{\label{subsec:spin_dyn} \NEW{Magnetic order}}

\begin{figure}[h]
\centering
\includegraphics[width=\linewidth]{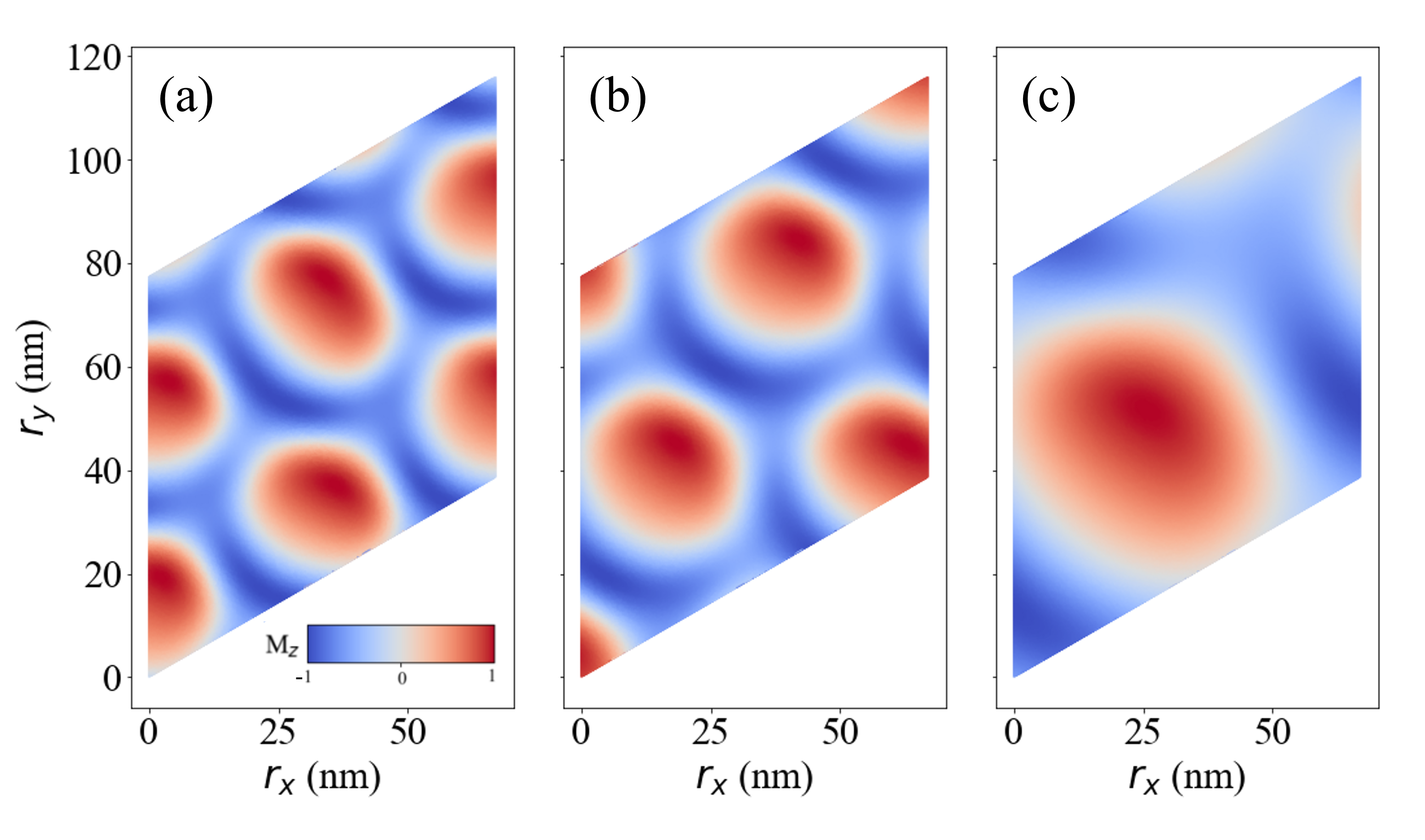}
\caption{\label{fig:sd_Ni} \small{Snapshot of a magnetic Ni layer configuration in: (a) Ni/Pt(111), (b) Ni/Ir$_{1}$/Pt(111) and (c) Ni/Ir$_{2}$/Pt(111), with $T=0$ K and $B=0$ T. The color represents the \textit{z}-component  of each spin (M$_{z}$).}}
\end{figure}

The extracted Heisenberg and DM interactions allow careful analyses of the magnetic order and finite temperature behaviours by means of atomistic spin dynamics. Firstly, we performed spin dynamics simulations for the Ni/Ir$_{n}$/Pt(111) systems, setting $B=0$ and $T=0$: the ground state configurations obtained were, almost degenerate, spin spirals and skyrmions (see 
\NEW{Sec.~\ref{subsec:finite-effects}).}
The energy difference of these two magnetic configurations is of the order of a few $\mu$eV/atom, what is below the estimated numerical accuracy of the present DFT calculations. We have then chosen to study the states which presented skyrmions (as shown in Fig.~\ref{fig:sd_Ni}), and considered the vector direction perpendicular to the ultrathin film as the \textit{z} axis. 
The skyrmions observed here are Néel-type (topological charge $q=-1$) and for Ni/Pt(111) and Ni/Ir$_{1}$/Pt(111), we also noticed the presence of metastable skyrmioniums (see \NEW{Sec.}~\ref{sec:skyrmionium}).
Spin dynamics calculations estimate a size of $\approx 15-20$ nm for both $n=0$ and $n=1$ cases, and $\approx 25-35$ nm for the $n=2$ system, in the absence of applied magnetic fields. Here, we note that the skyrmions' sizes can vary in the simulations.

We also note that while the skyrmions are nearly of the same size for $n=0$ and 1, the density decreases as the number of Ir layers increases. If the magnetic ground state would be a skyrmion crystal, all skyrmions would have the same size and a constant density. Thus, the fact that the skyrmions here  occur with  random variations in size and densities implies that these are, indeed, not the ground state solution. Instead, they occur as metastable states with very low activation energies.

As mentioned in 
\NEW{Sec.~\ref{sec:comp_methods}} 
the  spin  dynamics calculations have been performed using the Hamiltonian (Eq.~\ref{hamil}), where dipole-dipole interactions and magnetic anisotropies were neglected. In similar multilayered systems,  these values have been shown to be small \cite{Herve2018}. \NEW{Concerning the dipole-dipole interactions, the negligence is justified by the fact that Ni possesses a relatively small magnetic moment, compared, for instance, with Fe and Co, and only monolayer Ni films are considered here.} 
For systems where the DMI is the main mechanism for the formation of skyrmions, the skyrmions’ size is typically 5-100 nm \cite{Fert2017,Nagaosa2013}, \NEW{which fits the found skyrmions' diameters in the present work}. 
Regarding the magnetic anisotropy, since it is not clear from the literature if Ni/Pt(111) should present an in-plane or out-of-plane magnetic easy axis \cite{Bornemann2012,Krishnan1991,Wilhelm2000}, atomistic spin dynamics calculations considering uniaxial anisotropy $K$ with an out-of-plane easy axis were performed for the Ni/Ir$_{n}$/Pt(111) systems, with strengths of $K=0.2$ meV/atom ($n=0$) and $K=0.1$ meV/atom ($n=1,2$). In these tests, we found similar results as presented in the 
\NEW{Sec.~\ref{subsec:phase_diag},} 
however relatively altered (for instance, smaller magnetic fields are needed to stabilize the skyrmion lattices); metastable skyrmioniums are still obtained in the $n=0,1$ cases.

\subsection{\NEW{\label{subsec:finite-effects}Finite size effects}}

Ni/Ir$_{n}$/Pt(111) magnetic ground states  were determined by performing Monte Carlo and spin dynamics simulations using the UppASD code and comparing the spin spirals and skyrmion lattices energies. A spin spiral can be described by a wave vector $\textbf{q}$, through the following equation 

\begin{equation}
    \hat{e}(\textbf{r},\textbf{q}) = \hat{s}\cos(\textbf{r}\cdot \textbf{q}) + \hat{n}\sin(\textbf{r}\cdot \textbf{q})
\end{equation}

\noindent where $\hat{e}$ is the local magnetization direction, \textbf{r} is the spin position and $\hat{s}$ and $\hat{n}$ are unit vectors defining the rotational plane of the spiral. 
Given the two unit vectors, we can map out the energy landscape by calculating the total energies of all spin spirals which have  $\textbf{q}$-vectors  commensurate with the dimensions of the system. The spin spiral state with the $\textbf{q}$-vector corresponding to the lowest energy was then identified and relaxed further, using over-damped zero-temperature spin dynamics simulations to let the system relax as far as possible. Skyrmion lattices were simulated using the same approach but then $3\textbf{q}$ spin spirals, i.e. a magnetic texture described by 3 $\textbf{q}$-vectors with the same magnitude of $\textbf{q}$ but rotated 120$
^{\circ}$ between them, were considered. Furthermore, theses results were compared with the Monte Carlo (MC) simulations used to build the phase diagram.  Here, the influence of the lattice on the results was analyzed as well. Thus, the procedure described above was done for the proposed ultrathin films, considering a square lattice size ($N\times N$), with $200\leq N\leq 500$ for \textit{n} = 0, 1, 2. We also considered two types of $1\textbf{q}$ simulations: spin spirals in the $\hat{x}$ direction ($1\textbf{q}$ $yz$) and $\hat{y}$ direction ($1\textbf{q}$ $xz$). 

\begin{figure}[h]
\centering
\includegraphics[width=\linewidth]{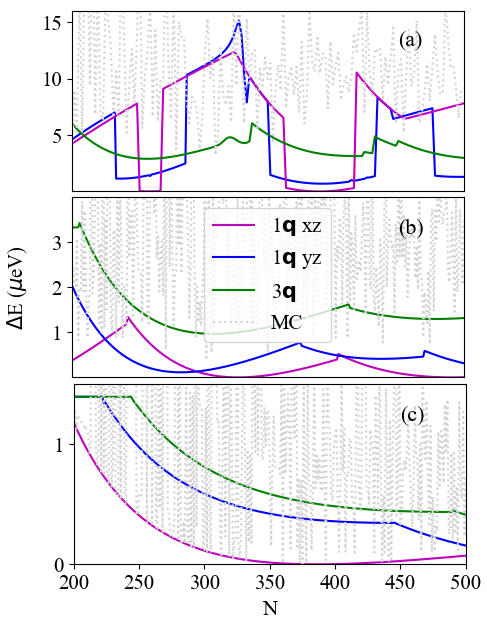}
\caption{\label{fig:ipQ} \small{Energy difference as a function of the lattice size N. Results for spins spirals (1\textbf{q}), skyrmions (3\textbf{q}) and Monte Carlo (MC) calculations performed for: (a) Ni/Pt(111); (b) Ni/Ir$_{1}$/Pt(111) and (c) Ni/Ir$_{2}$/Pt(111). The zero value is the minimal energy. Colour conventions are the same in all cases.}}
\end{figure}

The results for both $1\textbf{q}$ and $3\textbf{q}$ calculations for the Ni ultrathin films are shown in Fig.~\ref{fig:ipQ}. Notice that the energy vertical scale is in $\mu$eV, what implies that the curves are almost degenerate in the whole lattice size range studied here. 

Therefore, a typical lattice size of $N=280$ was chosen and, in this case,  for  \textit{n} = 0 and 1, the state with the minimum energy is a spin spiral in the $\hat{x}$ direction ($1\textbf{q}$ $yz$), while for $n=2$ the lower-energy state found is a spin spiral in the $\hat{y}$ direction ($1\textbf{q}$ $xz$). However, the energy difference between these configurations and skyrmions ($3\textbf{q}$) is, indeed, very low, indicating an almost degenerate  SS and SkL state. Considering $N=280$, our calculations show $\Delta E=$ 0.97 $\mu$eV for $n=0$, $\Delta E=0.96\,\mu$eV  for $n=1$, and $\Delta E=0.76\,\mu$eV for $n=2$, always favouring the SS phase. Furthermore, we notice that the hierarchy of the considered states, i.e. the   stability order  (total energy competition) clearly depends on the system size. This points to the necessity of always performing a careful finite size scaling analysis, when discussing the stability of complex magnetic states. 

As a result of the almost degeneracy obtained in the ground state, 
we can predict that spontaneous skyrmions shall be experimentally observed in these systems. We also note that, for \textit{n} = 2, the SkL range is less homogeneous than for \textit{n} = 0 and 1, intercalating the phase region with spin spiral state (\NEW{as presented in the} phase diagram, Fig.~\ref{fig:pd}(c)). This can be understood in terms of the small $\Delta E$ found (Fig.~\ref{fig:ipQ}), and less stable SkL under ${(B,T)}$ ranges  (Fig.~\ref{fig:pd}).

\begin{figure}[h]
\centering
\includegraphics[width=1.0\linewidth]{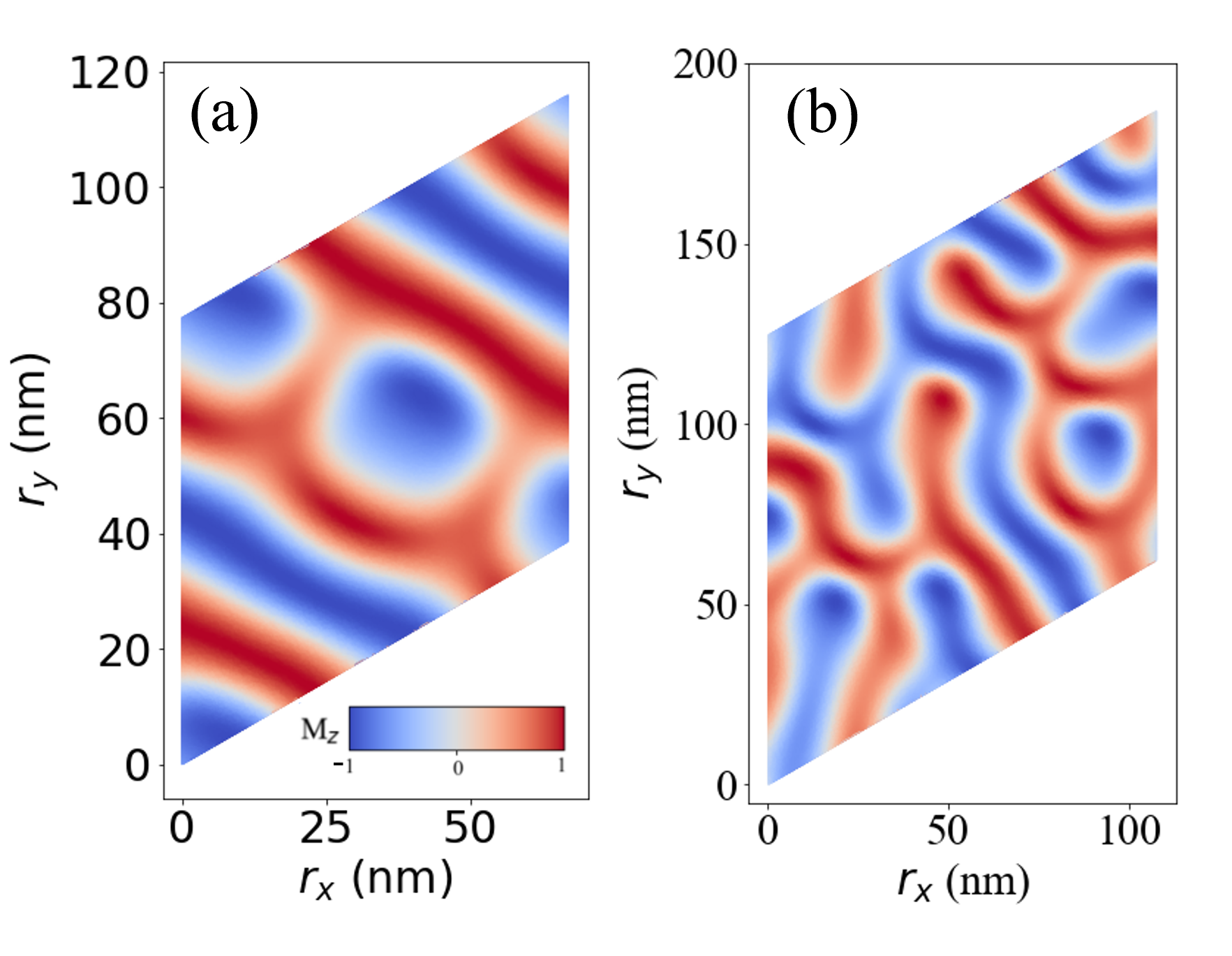}
\caption{\label{fig:n0_sizes} \small{Magnetic configuration, with $T=10^{-4}$ K and $B=0$ T, obtained with spin dynamics calculations for Ni/Pt(111) with spin lattice sizes of: (\textit{a}) $280\times280$  and (\textit{b}) $450\times450$.}}
\end{figure}

Since the energy difference between pure SS and skyrmionic states is small, here we also obtain in the ground state mixed states (skyrmions and spirals), where their sizes and density can vary in the simulations, as illustrated in Fig. \ref{fig:n0_sizes} for Ni/Pt(111).

\subsection{\NEW{Comparison with Co/Ir$_{n}$/Pt(111)}}
Further understanding of the physical mechanisms playing a role here can be achieved by analysing and comparing the behaviour of the Ni/Ir$_{n}$/Pt(111)  with other $3d$/Ir$_n$/Pt(111) systems. With this aim, we also performed calculations for Co/Ir$_n$/Pt(111), using the same methodology. Since this comparison is not the main focus of the present work,  details  for the Co based systems are shown in the Appendix~\ref{appenb:Co_systems}. The magnetic  Co/Ir$_{n}$/Pt(111) ground state configurations are composed by domain walls, according to the present results and in agreement with Ref. \cite{Vida2016}. This intriguingly  different ground state magnetic configurations for Co- and Ni-based systems can be explained by analyzing the Heisenberg exchange and DM interaction behaviours. The  Co-Co exchange interactions  are strongly ferromagnetic and larger than the  Ni-Ni, as could be expected. Nevertheless, the DM interaction in both, Co and Ni, systems are found to be of the same order of magnitude. Therefore,  the ratio of the $D_{1}$ component parallel to the plane ($D^{\parallel}_{1}$) to $J_{1}$ is larger  for Ni ultrathin layers compared to analogous Co-based systems (see Table~\ref{tab:ratio-JD} in the Appendix~\ref{appenb:Co_systems}). In addition, the DM vectors present different behaviours in Co/Ir$_{n}$/Pt(111) and Ni/Ir$_{n}$/Pt(111). While the $\vec{\textnormal{D}}_{1}^{\textnormal{Ni}}$ is mostly in the plane of the ultrathin film, $\vec{D}_{1}^{\textnormal{Co}}$ presents large out-of-plane components. Moreover, the sense of rotation of the Co-Co DM vectors changes as one includes Ir spacer layers (see Fig.~\ref{fig:dvec_Co} in  Appendix~\ref{appenb:Co_systems}). Therefore, we infer that the origin of the complex noncollinear magnetic nanostructures in the ground state of Ni/Ir$_{n}$/Pt(111) are related to the fact that $J_{ij}$ and $|\vec{D}_{ij}|$  have comparable magnitudes. Besides that, the presence of a significant $\vec{D}_{1}^{\textnormal{Ni}}$ component in the plane of the surface might also favour the formation of skyrmions in these systems.

\subsection{\label{subsec:phase_diag}Phase diagrams}

\begin{figure}[h]
\centering
\includegraphics[width=\linewidth]{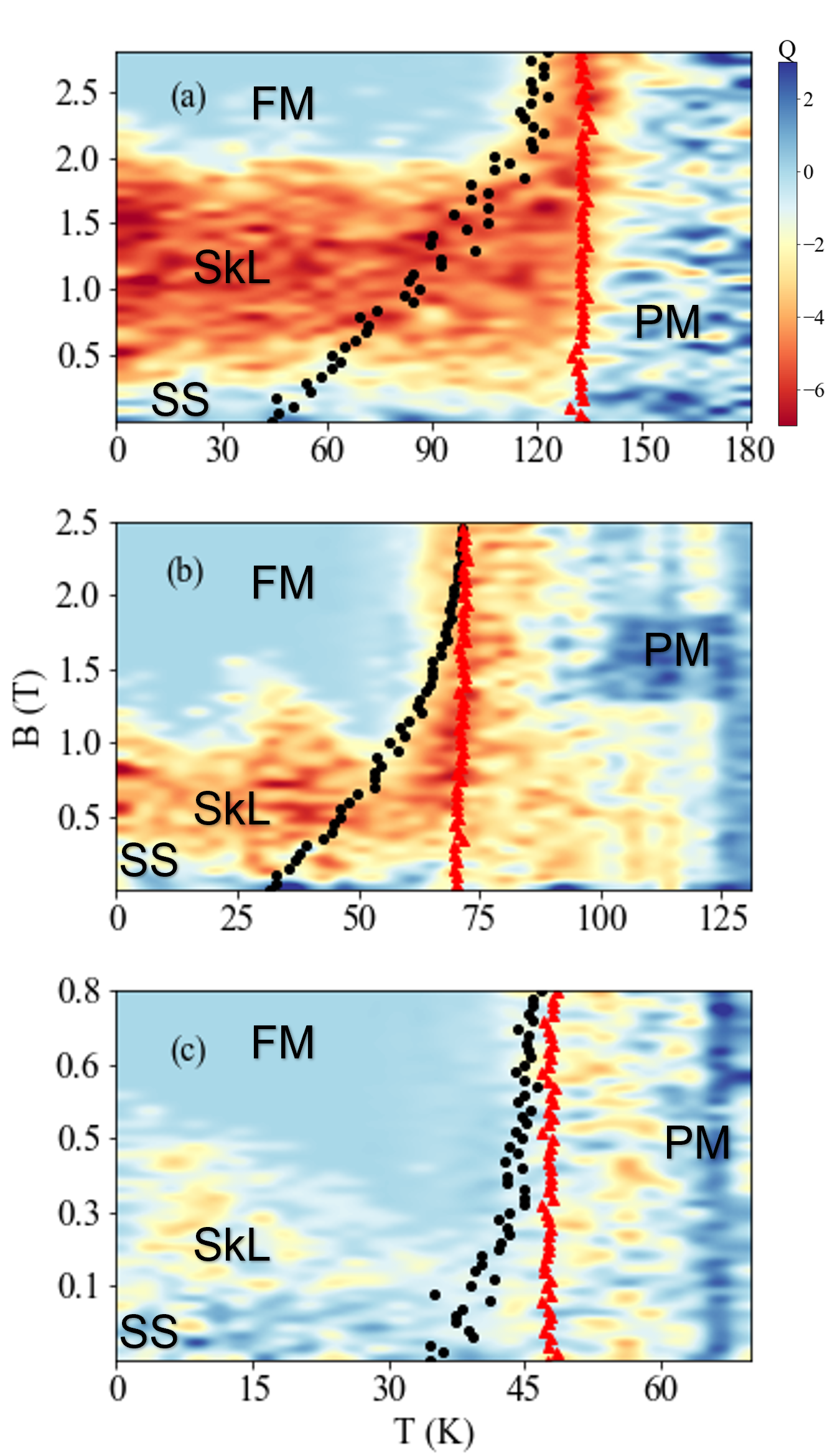}
\caption{\label{fig:pd} \small{Phase diagrams for (a) Ni/Pt(111), (b) Ni/Ir$_{1}$/Pt(111) and (c) Ni/Ir$_{2}$/Pt(111). Red (black) dots are the temperatures for which the spin susceptibility (specific heat) has a maximum value.
Colors code represent the topological charge of the lattice (Q). The ferromagnetic (FM), paramagnetic (PM), skyrmion lattice (SkL) and spin spirals (SS) phases are indicated in black.}}
\end{figure}

Now we turn our focus to the investigation of skyrmions stability with respect to temperature ($T$) and external magnetic field ($B$). In Fig.~\ref{fig:pd} we plot the time-average of the topological charge for our three systems. As discussed in \NEW{Sec.~\ref{subsec:finite-effects}}, we find that all systems have a spin-spiral (SS) minimum energy state with a minute energy difference with skyrmions at zero field and low temperatures, and a ferromagnetic single-domain (FM) order at large fields and low temperatures. Assuming that the topological charge is equal to the number of skyrmions in the system one would identify the orange areas in Fig.~\ref{fig:pd} as corresponding to a skyrmion lattice phase (SkL). However, it turns out that this is not a fully correct assumption because when analyzing the magnetic order visually, our simulations show that for a significant part of the phase space that is colored orange in Fig.~\ref{fig:pd}, the magnetic texture does not resemble a skyrmion lattice. Instead we find that at high temperatures the topological charge is built up by short lived and fluctuating point-like topological defects. For a given snapshot of the simulation, the topological charge is calculated to be finite but typically non-integer (i.e., does not correspond to actual skyrmions). Interestingly, despite the fluctuating nature of the point-defects, the time average of the topological charge is very similar to what is found for the ordered SkL. This behavior has also been seen in Pd/Fe/Ir(111) as reported in Ref.~\cite{Bottcher2018} where they refer to the region of point defects as an intermediate phase. 

According to the analysis above, the topological charge is seemingly not the most suitable order parameter to study phase transitions in these systems. In order to identify proper phase transitions, we thus also investigate the magnetic specific heat $c_V=\frac{dE}{dT}|_V$ which is expected to diverge at the critical temperature. The maximum values of the specific heat for every choice of external field is shown as red dots in Fig.~\ref{fig:pd}, obtained by a Lorentzian fitting. We find that the only phase transition that is clearly visible from an analysis of the specific heat is the order-disorder transition which occurs at $130$ K for the Ni/Pt(111) system. The critical temperature reduces when introducing Ir layers to $70$ K for Ni/Ir$_1$/Pt(111) and $47$ K for Ni/Ir$_2$/Pt(111). Based on the notion of the intermediate phase which is paramagnetic (PM) but has a finite topological charge (on average) and the specific heat analysis, one should be careful in interpreting the data shown in Fig.~\ref{fig:pd} as a true phase diagram. It does however fill the purpose of comparing the stability of the SkL between our three systems where we see that all systems have a spin-spiral ground state at zero field and low temperatures but where the area indicating stability of the skyrmion lattice decreases with increasing number of Ir layers. This is consistent to what was observed experimentally for the limiting case Ni/Ir(111) ($n\rightarrow\infty$), in which no magnetic contrast was found \cite{Iaia2016}. \NEW{Note that} this behaviour can be directly related to the decrease (Fig. \ref{fig:jDNi}(a)) on the $J_{ij}$ parameters. In addition to the specific heat, we also calculated the field-dependent spin susceptibility $\chi$ from the fluctuations of the magnetization as $\chi=\frac{1}{k_B T}(\langle {m^2}\rangle-{\langle{m}\rangle}^2)$, where $k_B$ is the Boltzmann constant. The resulting peaks for the susceptibility curves are shown as black dots in Fig.~\ref{fig:pd}.
Like the specific heat, the peaks of the calculated susceptibility are also expected to indicate phase transitions. We find that the susceptibility maxima do not agree with the specific heat peaks but coincides quite well with the visual inspection of the systems, i.e.  the susceptibility is a good indicator for when the SS or SkL states appears to be strongly disordered. We also find that the susceptibility maxima increases with increasing external field as expected.
 
 As mentioned in Sec.~\ref{subsec:abinitio}, we inspect the role of the inclusion or not of long range interactions in the spin dynamics simulations, as also addressed in Refs.~\cite{Leonov2015,Lin2016,Malottki2017}. For this we compared calculations including either only NN or 5 NN shells in the spin dynamics simulations for Ni/Pt(111). We verified that the lack of inclusion of further than NN neighbors affects the stability study of skyrmions in this case. Therefore the procedure used in the present work (including 5 NN shells) is indeed necessary.
 
\subsection{\label{sec:skyrmionium}Skyrmioniums}

\begin{figure}[h]
\centering
\includegraphics[width=1.0\linewidth]{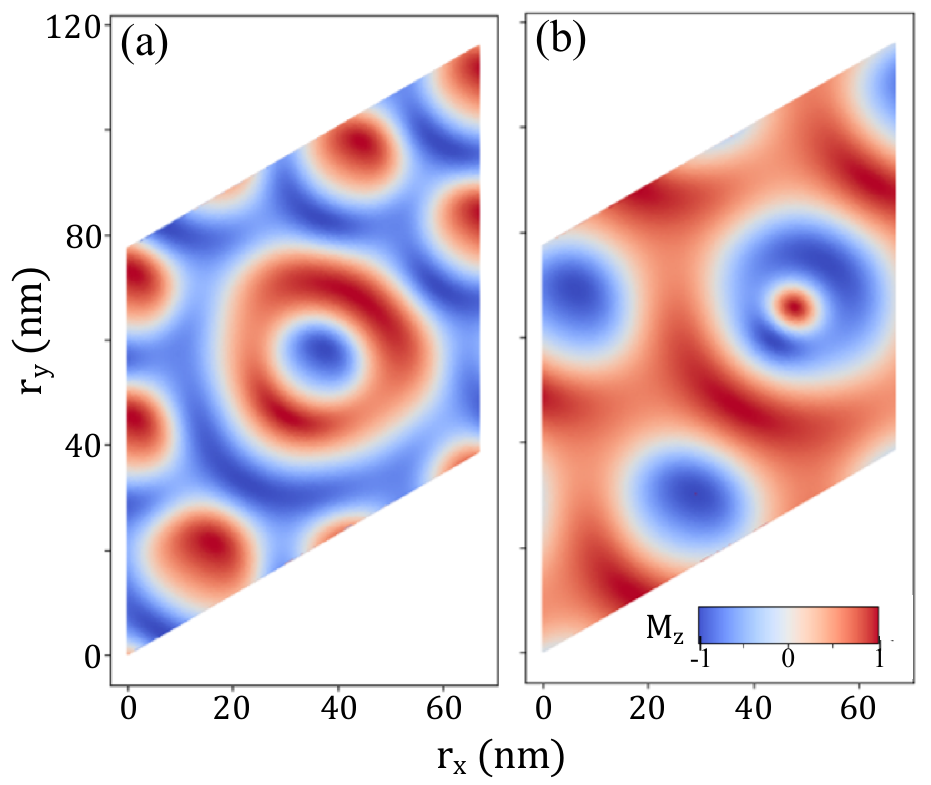}
\caption{\label{fig:skrmionium} \small{Magnetic configuration, with $T=10^{-4}$ K and $B=0$ T, obtained with spin dynamics calculations for: (\textit{a}) Ni/Pt(111) and (\textit{b}) Ni/Ir$_{1}$/Pt(111). Examples of states with spontaneous skyrmioniums. The color represents the \textit{z}-component  of each spin (M$_{z}$).}}
\end{figure}

 We found, for the Ni/Ir$_n$/Pt(111), spontaneous single skyrmioniums in the cases $n=0$ and $n=1$ with zero external magnetic field and low temperatures ($T=10^{-4}$ K):  examples of these states are shown in Fig.~\ref{fig:skrmionium}(a) and \NEW{Fig.~\ref{fig:skrmionium}}(b), respectively. The configurations presenting single skyrmioniums were found to be metastable with respect to the SS state 
 \NEW{(for details, see below)}
  and we expect that they can eventually appear in an experimental setup of Ni/Pt(111) and Ni/Ir$_1$/Pt(111).

\begin{figure}[h]
\centering
\includegraphics[width=1.0\linewidth]{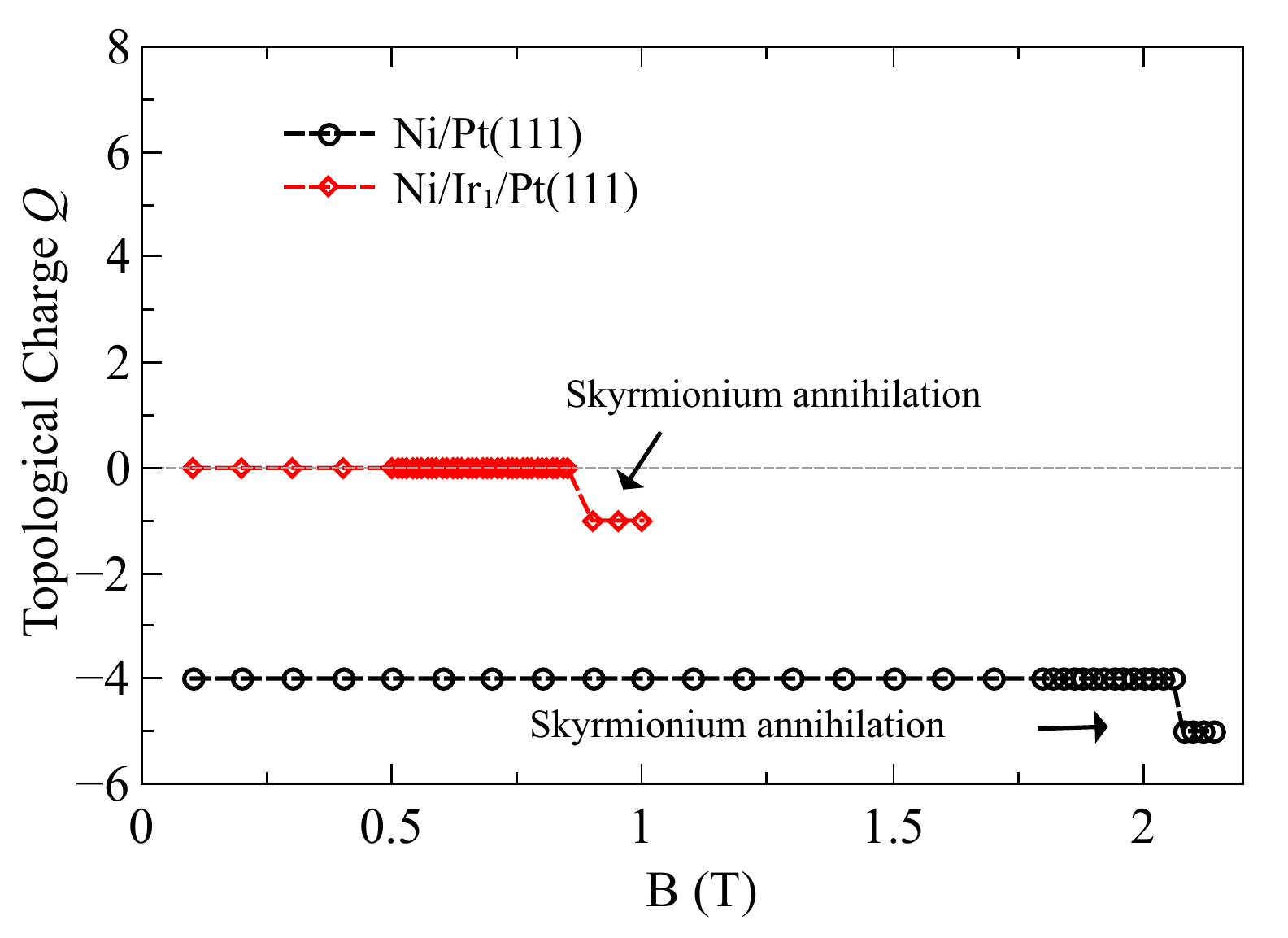}
\caption{\label{fig:q-b} \small{$Q-B$ diagrams for: (black) Ni/Pt(111) and (red) Ni/Ir$_{1}$/Pt(111). The arrows indicate the magnetic field ($B$) magnitude in which the skyrmioniums are annihilated. Results are for the low temperature regime of $T=10^{-4}$ K. The lines are guides to the eyes.}}
\end{figure}

We inspected the skyrmionium stability from two points of views:   external magnetic field and temperature. In the first case, we verified that the skyrmionium vanishes with $B\sim 2.08$ T for $n=0$ and $B\sim 0.9$ T for $n=1$. These limits could be extracted from the evolution of the topological charge of the system with respect to $B$ 
(see Fig.~\ref{fig:q-b}).
The lower critical value of $B$ for Ni/Ir$_1$/Pt(111) is expected, since from the phase diagrams the SkL occur up to a diminished maximum external field than for Ni/Pt(111). 
We note, however, that the critical field is slightly decreased in comparison to the obtained for skyrmions in the phase diagrams. 

\begin{figure}[h]
\centering
\includegraphics[width=1.0\linewidth]{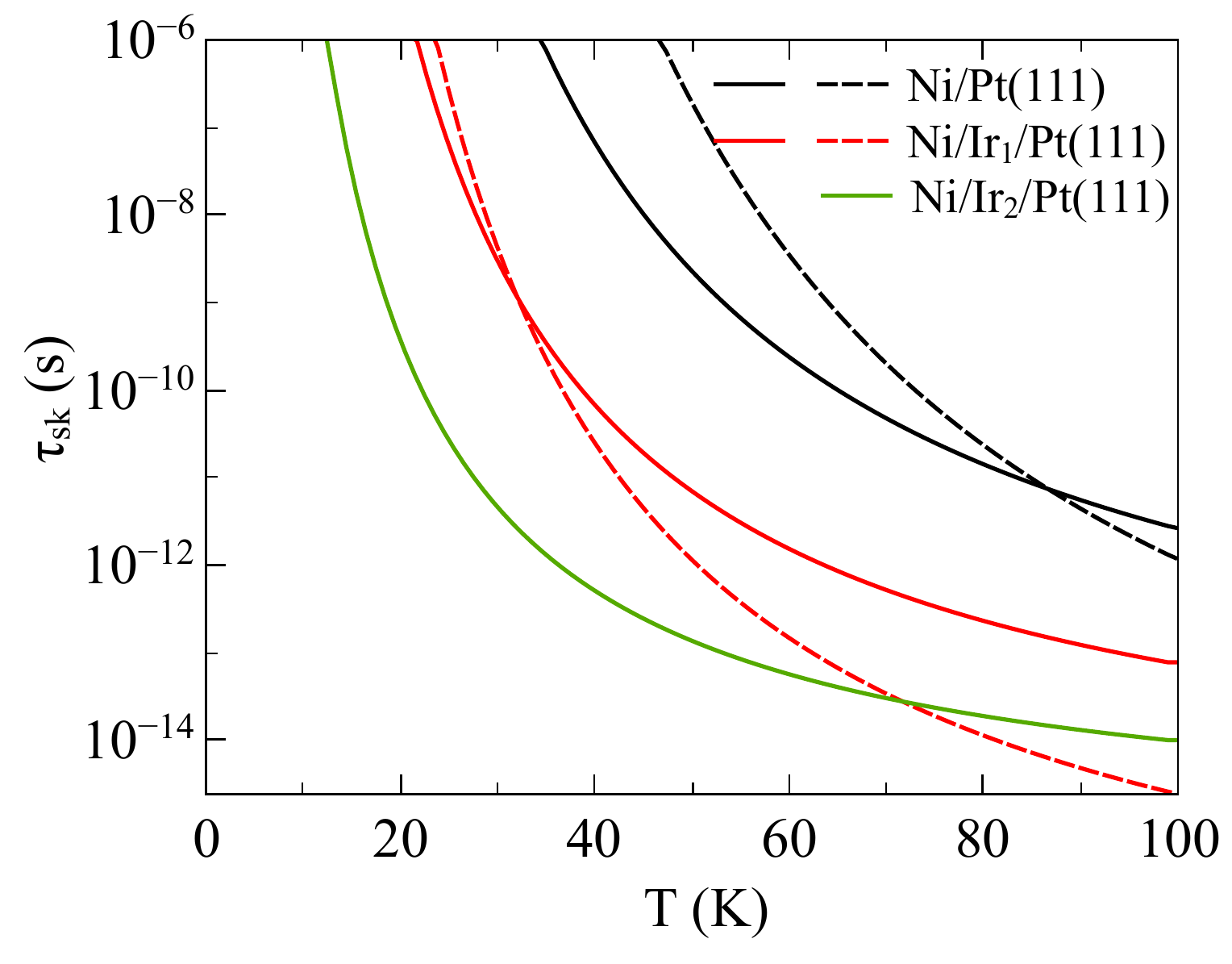}
\caption{\label{fig:stability-temperature} \small{Lifetime of skyrmions (full lines) and SkL + skyrmionium (dashed lines) configurations in Ni/Ir$_{n}$/Pt(111) as a function of temperature, considering $\textnormal{B}=0$ T, for configurations with skyrmioniums (Fig.~\ref{fig:skrmionium}).}}
\end{figure}

The influence of skyrmionium in the skyrmion lifetime ($\tau_{sk}$) as a function of the temperature, as well as the skyrmion lifetime itself, are shown in Fig.~\ref{fig:stability-temperature}. The calculation details are described 
\NEW{below.}
It is clear that the enhancement of the Ir buffer causes skyrmions to be more unstable from the temperature perspective. Moreover, our results demonstrate that the presence of skyrmioniums can non-negligibly affect $\tau_{sk}$, essentially due to the larger size of skyrmioniums and the consequent change in the entropy of the system \cite{Von2019}.

Since skyrmioniums are characterized by a topological charge $q=0$ (due to its doughnut-like structure composed by a skyrmion with $q=+1$ inside a skyrmion with $q=-1$, or vice-versa \cite{Zhang2016}) they are undetectable through $q$ charge measurements, as other non-topological spin textures (ferromagnetic and antiferromagnetic mediums, or spin-spirals). However, our calculations have shown that these structures can emerge together with skyrmions in Ni/Pt(111) and Ni/Ir$_{1}$/Pt(111) as metastable states, with an energy difference with respect to the SS state of $\Delta E=1.78$ $\mu$eV ($n=0$), and $\Delta E=3.27\,\mu$eV ($n=1$).

\begin{figure}[h]
\centering
\includegraphics[width=\linewidth]{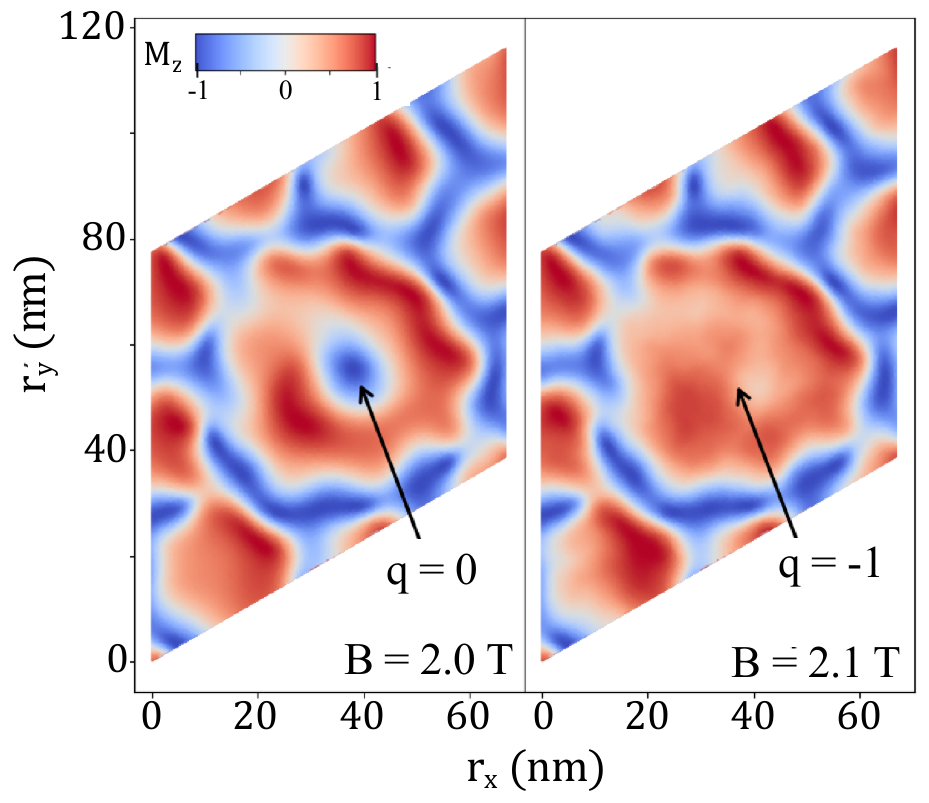}
\caption{\label{fig:sk-anni} \small{Skyrmionium annihilation process with the action of an external magnetic field from $B=2.0$ T to $B=2.1$ T in Ni/Pt(111), resulting in a topological  $q=0\rightarrow q=-1$ transition. Results for the low temperature regime of $T=10^{-4}$ K.}}
\end{figure}

Spin dynamics simulations reveal that the annihilation of skyrmioniums in these systems with the action of external magnetic fields $B$ occur by transforming them into regular skyrmions, what implies a topological  $q=0\rightarrow q=\pm1$ transition. Fig. \ref{fig:sk-anni}
exhibits an example of this transition for a skyrmionium in Ni/Pt(111), from $B=2.0$ T to $B=2.1$ T (and $T=10^{-4}$ K). 
Since we 
obtain the topological charge $Q$ of the whole spin system, the stability of the skyrmionium with respect to $B$ can be inferred from a $Q-B$ diagram, together with the visual inspection of the resulting spin state, for each analyzed magnetic field. The obtained $Q-B$ diagrams for Ni/Ir$_{n}$/Pt(111) (for $n=0$ and $n=1$) are shown in Fig.~\ref{fig:q-b}, for the specific configuration in Ni/Pt(111) shown in Fig. ~\ref{fig:sk-anni} and another particular configuration containing a skyrmionium for Ni/Ir$_{1}$/Pt(111). The magnetic fields were considered  up to the critical value for each system (see Fig. \ref{fig:pd}). We note that the skyrmionium annihilation events are instantaneous, analogously to what was obtained for regular skyrmions \cite{Rozsa2016}. We also remark that this field-stability of skyrmioniums should not depend on the damping $\alpha$, as it is intrinsically related to the interactions in the spin model.

From these results, we can determine the critical fields 
$B_{crit}\sim 2.08$ T for Ni/Pt(111) and $B_{crit}\sim 0.9$ T for Ni/Ir$_{1}$/Pt(111). As a consequence of these particular spin configurations, in which skyrmioniums can generate $q\neq 0$ structures, the topological charge fluctuates around $Q=0$, resulting in skyrmions even in a higher range of magnetic fields than the described in 
Fig. \ref{fig:pd}.

\begin{figure}[h]
\centering
\includegraphics[width=1.0\linewidth]{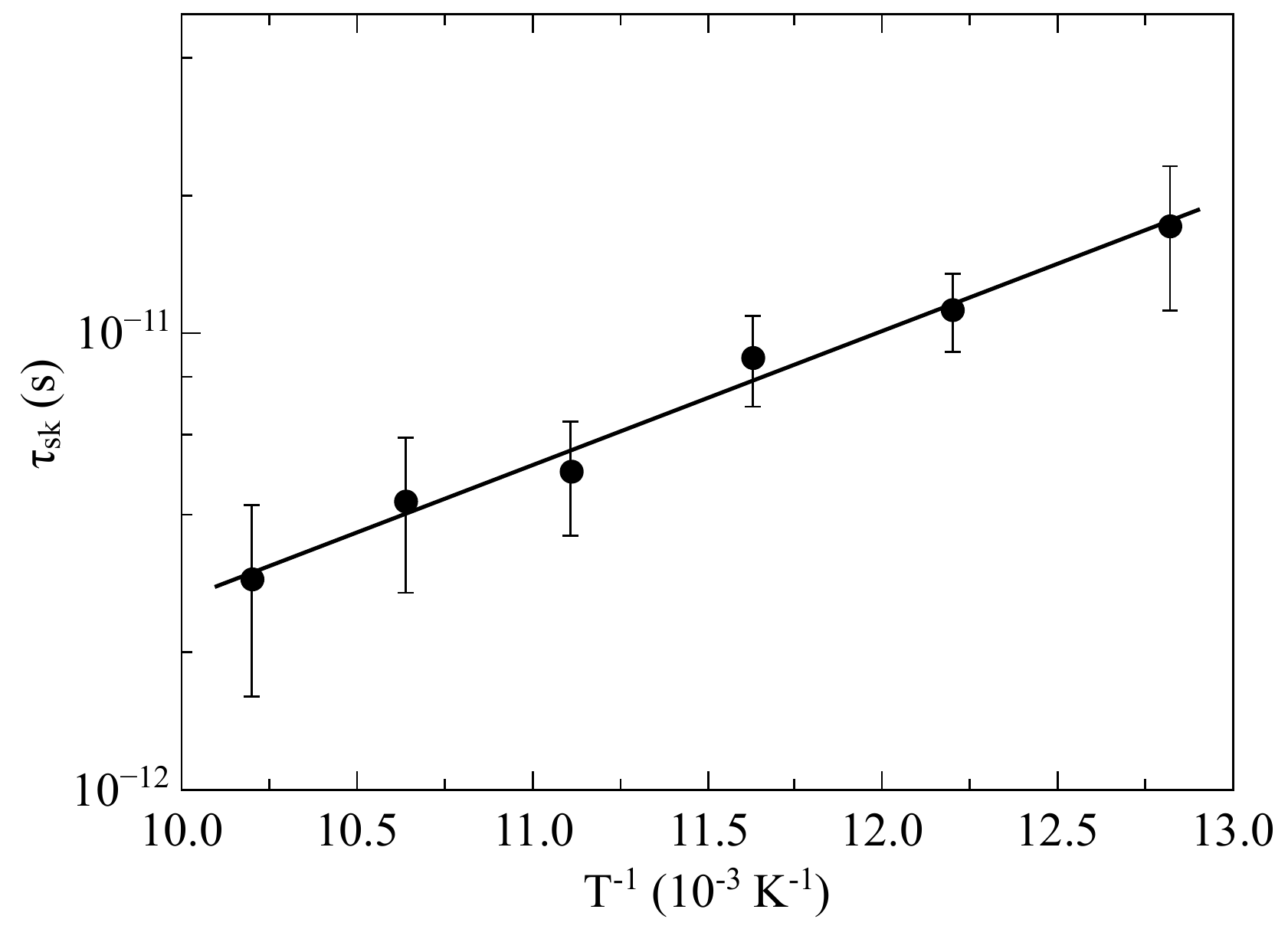}
\caption{\label{fig:tsk-nipt} \small{Skyrmion lifetime $\tau_{sk}$ as a function of the inverse of the temperature $T^{-1}$ for Ni/Pt(111), considering no external magnetic field ($B=0$).
Arrhenius law (Eq. \ref{eq:arrheinus}) fit (black line).
}}
\end{figure}

 We also investigated the influence of skyrmionium in skyrmion lifetime as a function of the temperature, and compared the results with the lifetime of the pristine skyrmion lattice ($\tau_{sk}$) in the same substrate. For this case, we considered $B=0$ in all studied configurations. According to previous works \cite{Von2019,Rozsa2016,Yin2016,Hagemeister2015,Wild2017}, it is known that the skyrmion creation is a nucleation process, and the lifetime should follow the Arrhenius law

\begin{equation}
\label{eq:arrheinus}
\tau_{sk}\sim\tau_{0}e^{\frac{\Delta E}{k_{B}T}},
\end{equation}

\noindent being $k_{B}$ and $\tau_{0}$, respectively, the Boltzmann constant and a proportionality constant (the so-called attempt time), and $\Delta E$ the collapse energy barrier. Therefore, it is expected that for skyrmioniums this also holds. As we calculate the skyrmion number of the whole spin system ($Q_{total}$), the lifetime can be extracted from the $Q(t)$ function considering that the increase in $Q_{total}$ is only due to the annihilation of downwards pointing skyrmions or the creation of a central-downwards pointing skyrmioniums (see Fig.~\ref{fig:sk-anni}, inverse process), the latter in the case of a system containing skyrmioniums. This means that the presence of upwards pointing skyrmions or antiskyrmions is excluded. In the stationary case (in which $\frac{dQ}{dt}=0$), the skyrmion lifetime is obtained by multiplying the average time between skyrmion annihilation events with the average topological charge, $\bar{Q}_{total}$, analogously to what was done in Ref.~\cite{Rozsa2016}. With this definition, we calculated $Q(t)$ in a time interval of 25 ps,
considering the Ni/Ir$_{n}$/Pt(111) SkL, and, in the case of $n=0$ and $n=1$, also the SkL with a skyrmionium configuration. An example of skyrmion lifetime as a function of temperature is shown in Fig.~\ref{fig:tsk-nipt}, for the Ni/Pt(111). By fitting Eq.~\ref{eq:arrheinus} in the $\tau_{sk}$ data, we obtain the collapse energy barriers and attempt times (here considered constant), which can be compared to the energy coming from the largest contributions of isotropic exchange interactions, from first to third neighbours (in the number of six each). Skyrmions are unwound if the central spin is rotated to the direction of the most external spin, and the strongest contributions for the stabilization of this central spin come from the nearest exchange interactions. Therefore, with $\Delta E$ and $\tau_{0}$ obtained, then $\tau_{sk}(t)$ is determined (see Fig.~\ref{fig:stability-temperature}). The calculated collapse barriers decrease with larger thickness of the Ir buffer.

Similarly, we can consider the metastable configurations containing skyrmioniums for $n=0$ and $n=1$, shown in Fig.~\ref{fig:skrmionium}.
The same analysis is performed as in the pure SkL case, but now with the influence of these $q=0$ structures. 
We notice an enhancement of the collapse energy barriers in both surfaces, what is consistent with von Malottki \textit{et al.} \cite{Von2019} findings for Pd/Fe bilayers on Ir(111) and on Rh(111), where $\Delta E$ enhances with the skyrmion size. The skyrmioniums  radius are larger than in skyrmions, and contributes to the enhancement of $\Delta E$ in the entire magnetic configuration. This increase in $\Delta E$ diminishes the attempt time $\tau_{0}$ by the compensation effect, related to entropy, reported by Wild \textit{et al.} \cite{Wild2017} for the B20 Fe$_{0.5}$Co$_{0.5}$Si bulk.


\subsection{\label{subsec:speed}Spin-orbit torque driven dynamics}

Given the small relative thickness of Ni (and, in consequence, its high resistivity \cite{Sondheimer2001}) in comparison with the nonmagnetic substrate in Ni/Ir$_n$/Pt(111), it is assumed that the majority of an applied current density $\vec{j}$ will flow through Ir and/or Pt \cite{Ritzmann2018}. The spin Hall effect (SHE) of the heavy metal layers (Ir and/or Pt) converts the applied charge current along them in a perpendicular spin current injected in the Ni layer \cite{Sampaio2013}.  Therefore, the main mechanism expected to drive the skyrmions and skyrmioniums in the studied systems is the so-called spin-orbit torque (SOT) \cite{Emori2013,Manchon2019}.

In order to investigate the velocity of skyrmions and skyrmioniums in the Ni/Ir$_n$/Pt(111) systems, we followed the model of Eq.~\ref{eq:llg}, implemented in UppASD. In the SOT-induced dynamics, one can expect, under the assumption of a steady state motion, a linear relation between $j$ and the velocity $v$ ($v\propto j$)  \cite{Lee2018}. This also holds for skyrmioniums \cite{Zhang2016}. A suggestion of reading speeds estimation is given in \NEW{Sec. \ref{sec:Reading-speed}}.

\begin{figure}[h]
\centering
\includegraphics[width=\linewidth]{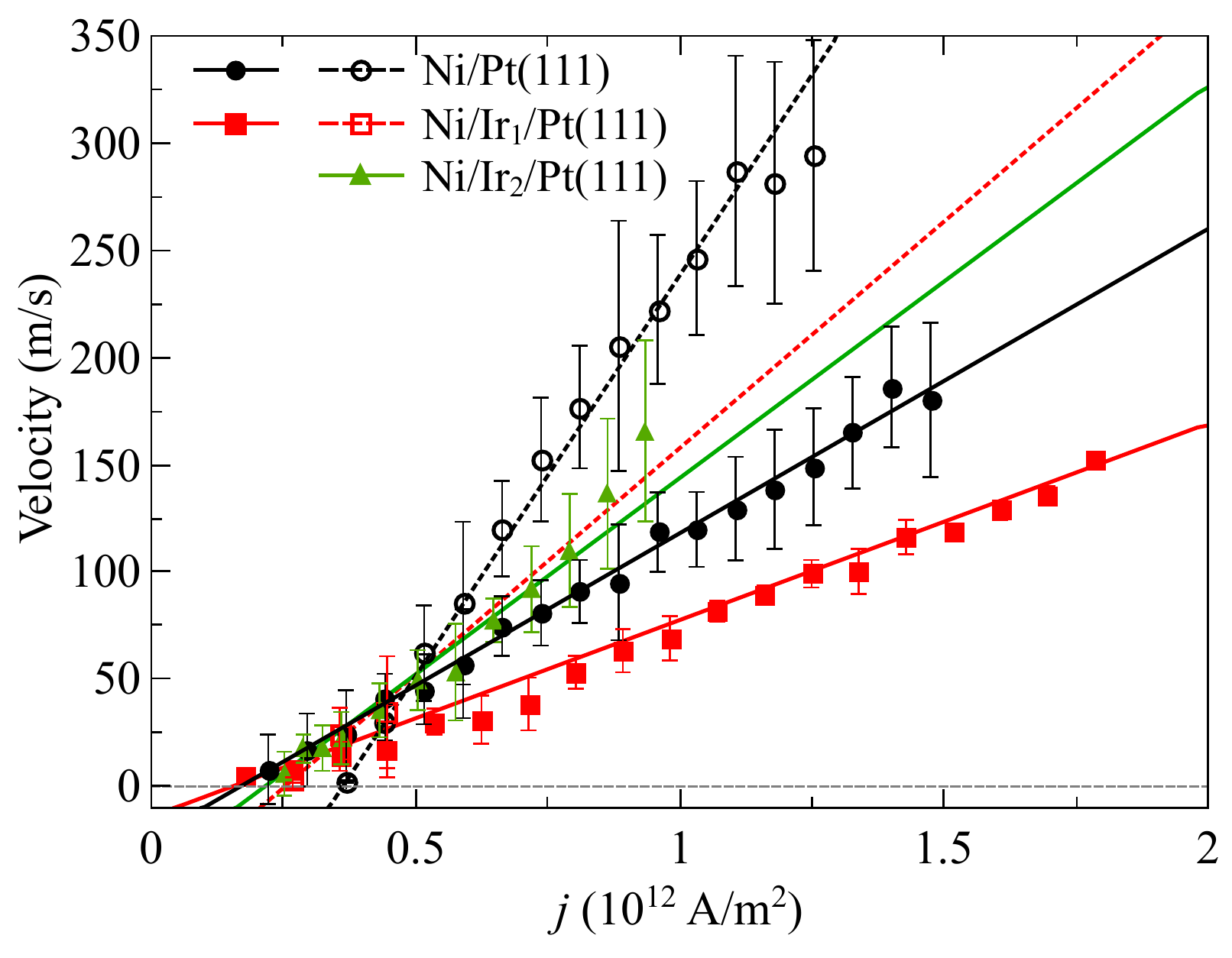}
\caption{\label{fig:speed} \small{Simulated skyrmion (solid points and lines) and skyrmionium (empty points and dashed lines) SOT velocity for Ni/Ir$_{n}$/Pt(111) as a function of the applied current density ($j$) in the $[\bar{1}10]$ direction (in-plane). In the dynamics, $T=10^{-4}$ K, and distinct $B$ were considered: $B=1$ T ($n=0,1$) and $B=0.5$ T ($n=2$) for skyrmions; and $B=0$ T for skyrmioniums.} The lines represent the fit of the linear function $v(j)=\kappa j+C$.}
\end{figure}

Our results of skyrmions and skyrmioniums motion in Ni/Ir$_n$/Pt(111) are shown in Fig.~\ref{fig:speed}, in which a linear function of the form $v(j)=\kappa j+C$ is fitted ($C$ is a constant, and the slope $\kappa$ is the current-driven mobility  \cite{Deger2019,Buttner2018}). With the aim of analyzing the velocity in a proper SkL configuration, we consider $B=1$ T for $n=0,1$, and, as in this magnitude the magnetization of Ni/Ir$_2$/Pt(111) is fully saturated, we choose the maximum possible field for the SkL phase in this system, $B=0.5$ T (see Fig.~\ref{fig:pd}). On the other hand, as spontaneous skyrmioniums are metastable for $n=0,1$, their dynamics is calculated at zero external field. The error bars were calculated considering the standard deviation of all velocities obtained from different skyrmions in the same system/configuration, and subjected by the same $j$. From the figure we notice the linear dependence of the velocity as a function of the current density. 
In turn, by inspection of the slopes, we observe that the inclusion one Ir layer between Ni and Pt(111) makes the skyrmions/skyrmioniums slower for a given $j$, while the inclusion of two Ir layers induces a higher current-driven mobility. This can be related to the damping parameter, larger in the $n=1$ case, given that $v\propto\alpha^{-1}$ \cite{Juge2019}.  Furthermore, with smaller saturation magnetization, Ni/Ir$_2$/Pt(111) SOT dynamics experience a stronger damping-like prefactor. At very low fields, skyrmions in Ni/Ir$_2$/Pt(111), larger than in Ni/Pt(111) and Ni/Ir$_1$/Pt(111), are more deformed via SOT. This tends to enhance even more their velocity, which can be explained in terms of the Thiele equation approach \cite{Zhang2016,Gobel2019,Thiele1973,Sampaio2013}. Regarding skyrmionium dynamics, we find higher $\kappa$ values when compared to skyrmions in the same substrate, which agrees with previous findings for different systems \cite{Gobel2019,Shen2018,Zhang2016}. It is related to the larger size of skyrmioniums and a consequently stronger deformation \cite{Gobel2019}. With the linear fit of the obtained data, we can extract the critical current density ($j_{c}$) for each system as $j_{c}=-\frac{C}{\kappa}$, which do not depend on $\alpha$. For skyrmion motion, we find similar values between $n=0,1$ systems ($0.17\pm0.02$ and $0.15\pm0.02$ TA$\cdot$m$^{-2}$, respectively), and slightly larger for $n=2$ ($0.22\pm0.03$ TA$\cdot$m$^{-2}$). For skyrmioniums, both $j_{c}$ values are enhanced compared to the skyrmion cases: $0.36\pm0.01$ TA$\cdot$m$^{-2}$ in Ni/Pt(111) and $0.25\pm0.05$ TA$\cdot$m$^{-2}$ in Ni/Ir$_1$/Pt(111).

\begin{figure}[h]
\centering
\includegraphics[width=\linewidth]{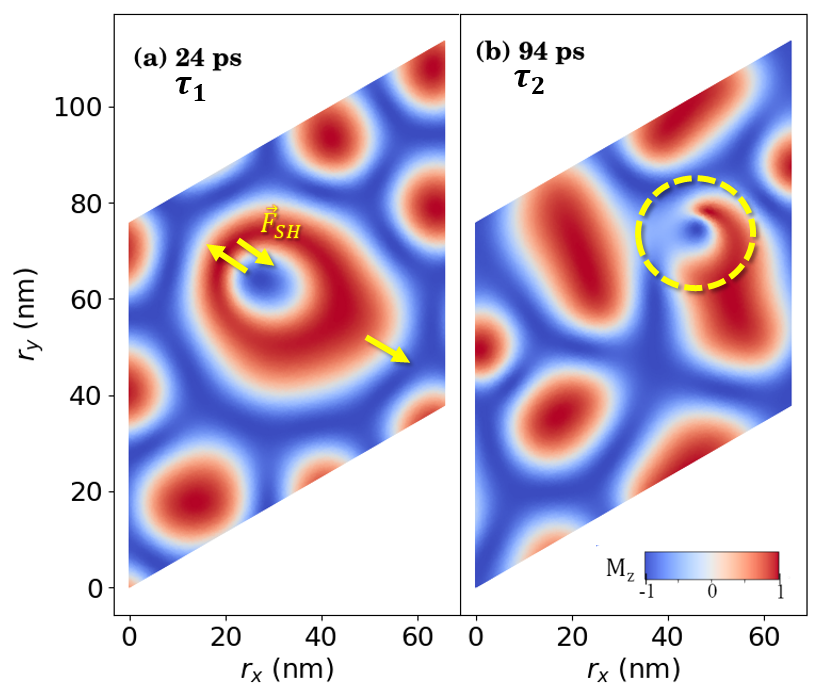}
\caption{\label{fig:unzipping} \small{Unzipping of a spontaneous skyrmionium (with $T=10^{-4}$ K and $B=0$ T) in Ni/Pt(111) for a current density of $j\sim1.47$ TA$\cdot$m$^{-2}$ in the $[\bar{1}10]$ direction. Calculated time interval of $\Delta\tau=70$ ps. $\vec{F}_{SH}$ is the spin Hall force. The color represents the $z$-component of each spin (M$_z$).}}
\end{figure}

For high current densities in both Ni/Pt(111) and Ni/Ir$_1$/Pt(111), we obtained an unzipping process of skyrmioniums \cite{Zhang2016,Xia2020}. This large $j$ enhances the transverse opposite driving spin Hall forces ($\vec{F}_{SH}$) from SkHE in the inner and outer skyrmions in a way that unbalances the confinement provided by the skyrmionium ring \cite{Gobel2020}, compelling the structure to suffer a massive distortion and a consequent annihilation. In these conditions, the skyrmionium is transformed into to two separate spin textures with $q=\pm 1$. In the simulation time interval, this skyrmionium annihilation is observed in the critical currents of $j_{c}\sim1.40$ TA$\cdot$m$^{-2}$ for Ni/Pt(111) and $j_{c}\sim0.54$ TA$\cdot$m$^{-2}$ for Ni/Ir$_1$/Pt(111). Fig.~\ref{fig:unzipping} shows an example of this process for the Ni/Pt(111) system and considering $j\sim 1.47$ TA$\cdot$m$^{-2}$, between the time instants of $\tau_{1}=24$ ps and $\tau_{2}=94$ ps from the beginning of the simulation (time interval $\Delta\tau=70$ ps). The deformation of the skyrmionium also causes the distortion of the skyrmions around in time. However, for much lower current densities, the transverse forces are weak enough to be cancelled by the skyrmionium ring confinement (which impels an opposite force), and the structure is maintained along the track.

\subsection{\label{sec:Reading-speed} Reading speeds}

\begin{figure}[h]
\centering
\includegraphics[width=\linewidth]{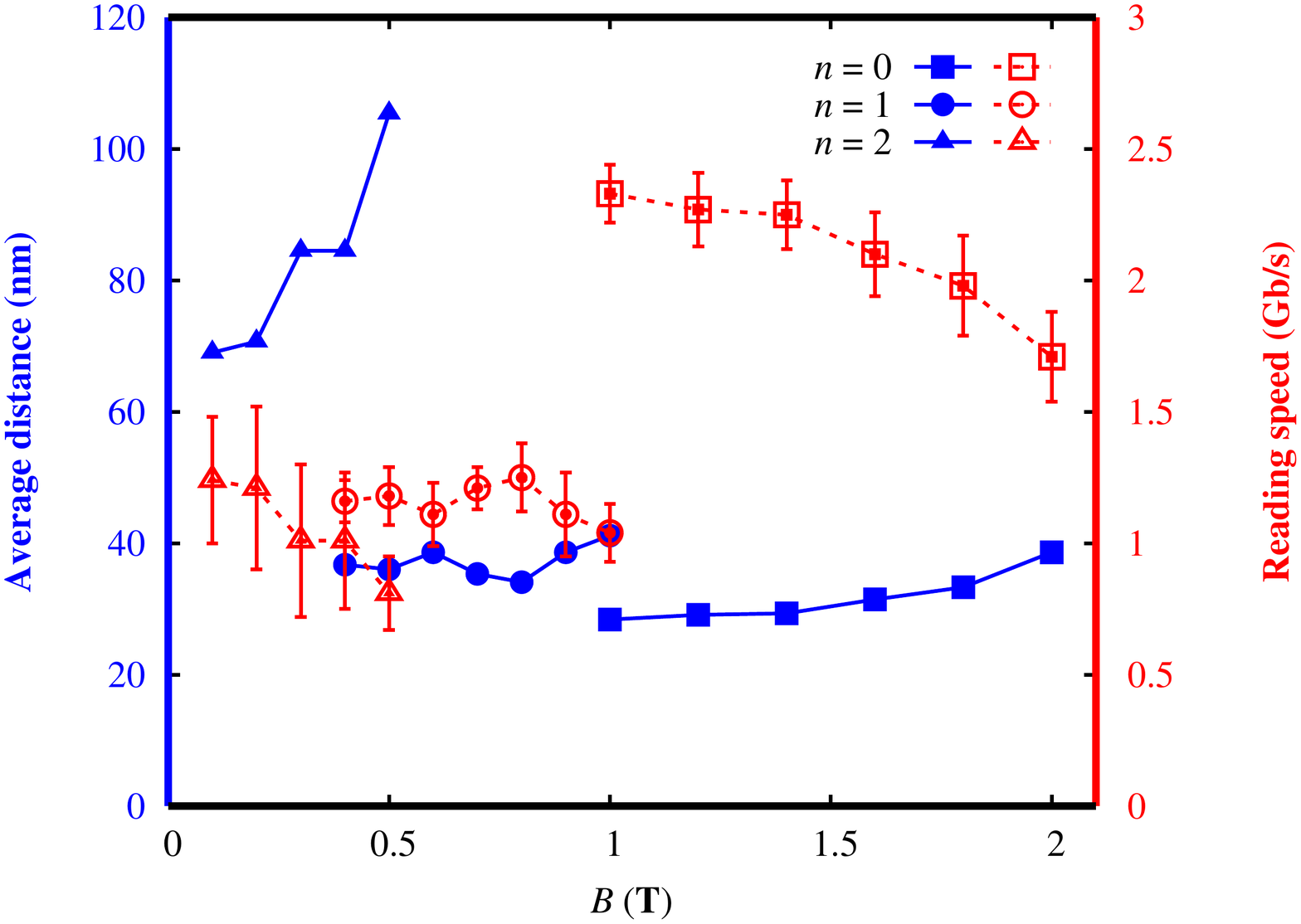}
\caption{\label{fig:reading_dist} \small{Reading speeds (in red, open symbols) and average distance between skyrmions (in blue, full symbols) in the assumption of a single racetrack memory composed by: Ni/Pt(111) (squares); Ni/Ir$_1$/Pt(111) (circles); and Ni/Ir$_2$/Pt(111) (triangles). In all simulations, we consider $T=10^{-4}$ K.}}
\end{figure}

In the assumption of a single racetrack memory \cite{Fert2013,Tomasello2014} composed by the Ni/Ir$_{n}$/Pt(111) systems, the expected reading speed can be obtained as the ratio of the velocity $v$ at a given current density and the average distance between each skyrmion \cite{Deger2019,Wu2017}. In this hypothesis, it is also straightforward to consider that each skyrmion represents a bit of information. Therefore, the calculated reading speeds (in Gb/s) of Ni/Ir$_{n}$/Pt(111), as well as the average separation between each skyrmion (in nm) for the external field ranges in which a SkL is found (see Fig. \ref{fig:pd}), are exhibited in Fig. \ref{fig:reading_dist}. Here, we always consider a constant current density of $j\sim0.5$ TA$\cdot$m$^{-2}$. As recently shown \cite{Litzius2020}, the velocity depends strongly on $T$, so it is expected that those rates will vary for different temperatures.

\section{\label{sec:conclusions}CONCLUSIONS}

  Local magnetic properties and the lower-energy state configurations of Ni/Ir$_{n}$/Pt(111) films (with $n=0,1,2$) have been here obtained using first principles calculations and spin dynamics. The minimum energy states are
  predicted to be spin-spirals but with a low energy difference with skyrmionic states.
  The presence of metastable spontaneous skyrmions in Ni/Ir$_{n}$/Pt(111), with low activation energies, points to a different behavior, when compared to other similar metallic ultrathin layers. We infer that this is related to the fact that 
the  $D_{1}$ to $J_{1}$ ratio is larger compared to Co/Ir$_{n}$/Pt(111), and that no significant out-of-plane DM vector component was obtained for the Ni-based systems. 
The skyrmions found here are small (diameters ranging from $\sim15-35$ nm), but with the enhancement of the Ir buffer thickness they become: larger, more dispersed and with more restricted $(B,T)$ conditions to occur, as well as their lifetimes decay. Regarding their velocity trends, the inclusion of one Ir layer between Ni and Pt makes the skyrmion slower for a given current density, while two Ir layers induce the higher current-driven mobility among the cases studied.
For $n=0$ and $n=1$ metastable skyrmioniums were also found which, compared to the skyrmions, present enlarged critical current densities and slightly lower stabilities with respect to the external magnetic field. Due to their larger size, this can enhance the lifetime of the magnetic configuration when compared to a clean (pristine) SkL and have shown to present a higher current-driven mobility than skyrmions. Therefore, this research not only contributes to the fundamental understanding of skyrmions and skyrmioniums in ultrathin films, but also opens a route towards the control of their properties via buffer thickness.

\section{\label{sec:acknowledgments}ACKNOWLEDGMENTS}

H.M.P. and A.B.K. acknowledge financial
support from CAPES, CNPq and FAPESP, Brazil. A.B. acknowledges eSSENCE. 
I.M. acknowledges financial support from CAPES, Finance Code 001, process n$^{\circ}$ 88882.332894/2018-01, and in the Institutional Program of Overseas Sandwich Doctorate, process n$^{\circ}$ 88881.187258/2018-01. P.C.C. acknowledges financial support from FAPESP, process 2020/05609-7.
The calculations were performed at the computational
facilities of the HPC-USP/CENAPAD-UNICAMP (Brazil), at the National Laboratory for Scientific Computing (LNCC/MCTI, Brazil), and at the Swedish National Infrastructure for Computing (SNIC).

\appendix

\section{\label{sec:appena:Ni_es} Ni/Ir$_n$/Pt(111) electronic structure}

\begin{figure}[h]
\centering
\includegraphics[width=\linewidth]{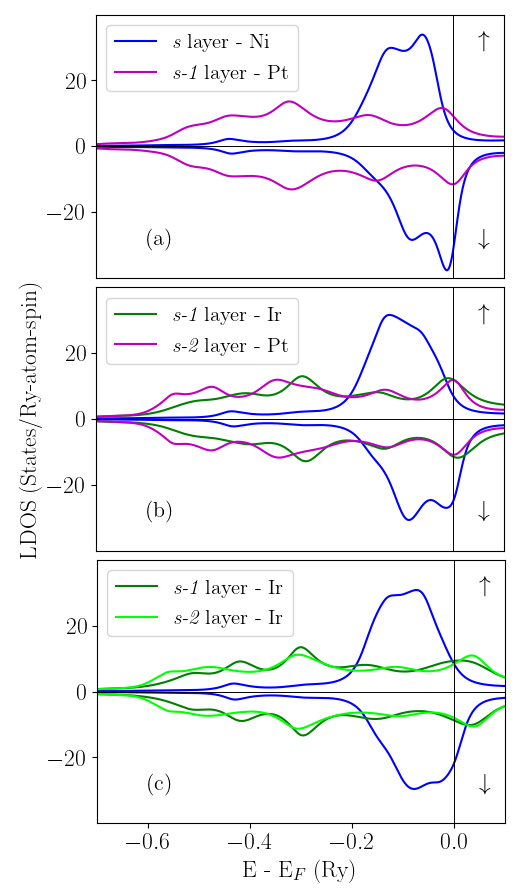}
\caption{\label{fig:ldos_Ni} \small{Spin polarized LDOS at a typical atom in different layers (\textit{s}, \textit{s-1} or \textit{s-2}) in (a)
Ni/Pt(111); (b) Ni/Ir$_{1}$/Pt(111) and (c) Ni/Ir$_{2}$/Pt(111). Color conventions are: Ni (blue), Pt (pink) and Ir (light and dark green).
}}
\end{figure}

The Local Density of States (LDOS)  projected on a typical atom in the \textit{s}, \textit{s-1} and \textit{s-2} layers, for the Ni/Ir$_{n}$/Pt(111) systems studied here, are shown in Fig.~\ref{fig:ldos_Ni}, where
a charge transfer between 3\textit{d} (Ni) and 5\textit{d} (Ir or Pt) orbitals can be observed.
After adding one Ir layer (\textit{n} = 1), the Ni electronic structure is changed and  3\textit{d} electrons, from both majority and minority bands, are transferred to Ir 5\textit{d} orbitals, leading to a decrease in the ${m}_{\textnormal{Ni}}$ and inducing a non-null magnetic moment in the \textit{s-1} layer. The induced magnetic moments in the  Ni/Pt(111) and Ni/Ir$_{1}$/Pt(111) \textit{s-1} layers are the same, indicating that the Ir and Pt layers exhibit analogous  behaviours. For \textit{n} = 2, the charge transfer is still responsible for the ${m}_{\textnormal{Ni}}$ reduction, but the Ir electronic structure is changed (in both layers \textit{s-1} and \textit{s-2}) and  its the majority and minority bands are almost symmetric. 
Therefore, the addition of two Ir layers (\textit{n} = 2) causes the induced magnetic moment in the \textit{s-1} layer to decrease about 50$\%$ with respect to \textit{n} = 0. 
We notice that, since the \textit{s} and \textit{p}  LDOS are relatively symmetric in all cases, we limit our analysis to the \textit{d} electrons.

\section{\label{appenb:Co_systems}Co/Ir$_{n}$/Pt(111) electronic structure}

Using the RS-LMTO-ASA code, and an analogous procedure (same cluster size and lattice parameters as described for Ni/Ir$_{n}$/Pt(111)) but replacing Ni by Co at the \textit{s} layer, we investigated Co/Ir$_{n}$/Pt(111) ultrathin films with \textit{n} = 0, 1 and 2.

As shown in Table~\ref{tab:mommag_co}
the Co local magnetic moment ($m_{\textnormal{Co}}$) is not very sensitive to the Ir spacer thickness but the induced magnetic moment in the \textit{s-1} layers decreases with increasing \textit{n}.

\begin{table}[h]
	\centering
	\caption{\small{Local magnetic moment ${m}$ (in $\mu_{B}$/atom) of a typical atom in the first three layers of the Co/Ir$_{n}$/Pt(111) systems.  The layers are denoted by \textit{s}, \textit{s-1} and \textit{s-2} (see text).}}
\begin{ruledtabular}
	\begin{tabular}{c|cc|cc|cc}
		& \multicolumn{2}{c|}{\textit{n} = 0} & \multicolumn{2}{c|}{\textit{n} = 1} & \multicolumn{2}{c}{\textit{n} = 2} \\ \hline
	\textit{s} &	Co & 1.80 & Co & 1.85 & Co & 1.82 \\
	\textit{s-1} & Pt & 0.31 & Ir & 0.28  & Ir & 0.19  \\
	\textit{s-2} & Pt & 0.14 & Pt & 0.05  & Ir & -0.04  \\
	\end{tabular}
	\label{tab:mommag_co}
\end{ruledtabular}
\end{table}

\begin{figure}[h]
\centering
\includegraphics[width=\linewidth, height=11cm]{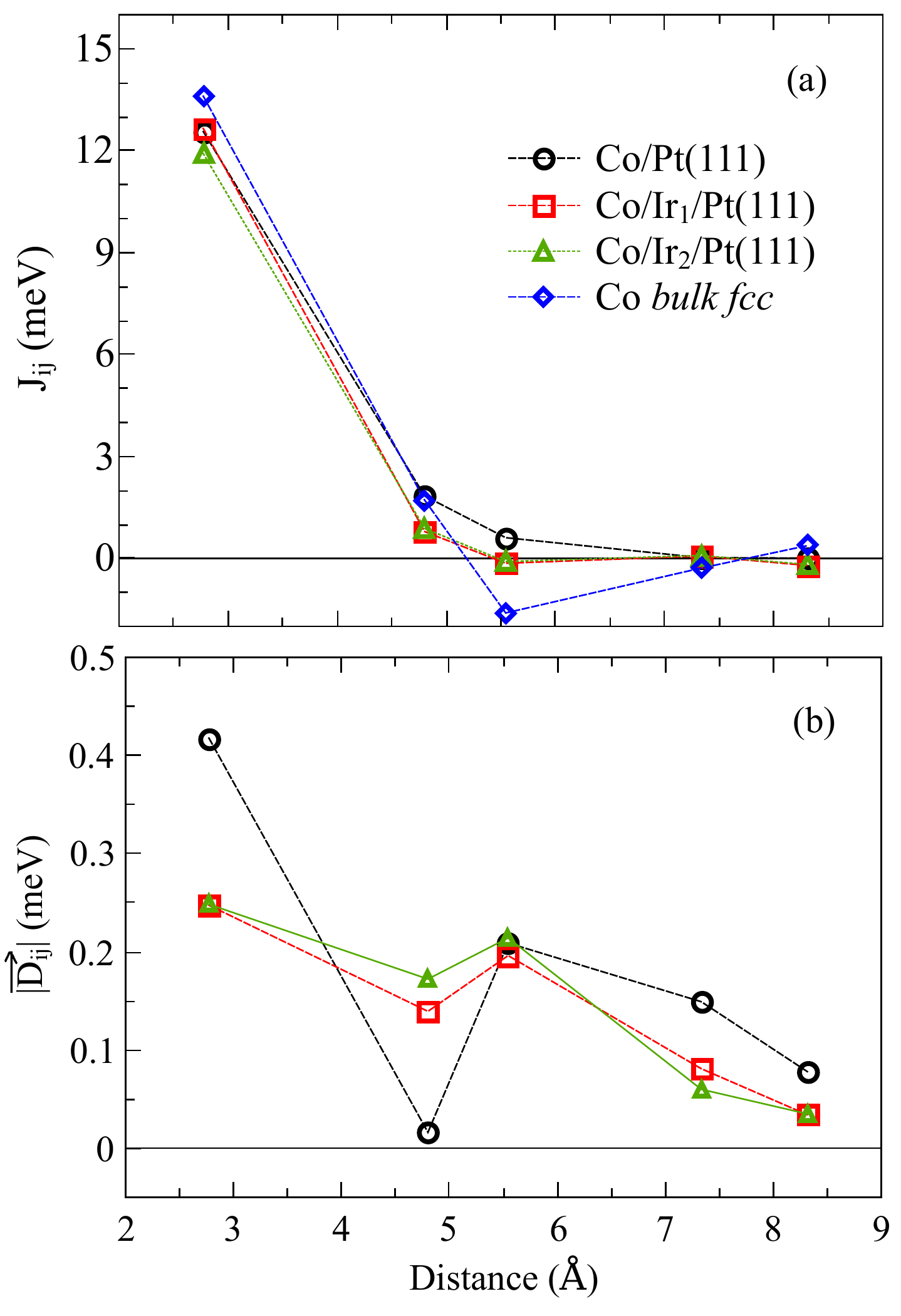}
\caption{\small{Co-Co (a) Heisenberg exchange coupling 
($J_{ij}$) and (b) DM interaction strength ($|\vec{D}_{ij}|$) as a function of the interatomic distance in Co/Ir$_{n}$/Pt(111) for \textit{n} = 0, 1 e 2.  Data for Co \textit{bulk fcc} obtained from Ref. \cite{Sonia2000}. The dotted lines are guides to the eyes.}
\label{fig:jDCo}}
\end{figure}

 Heisenberg exchange and DM interactions  for Co/Ir$_{n}$/Pt(111) are presented in Fig.~\ref{fig:jDCo}; also shown $J_{ij}$ data for Co \textit{bulk fcc}  from Ref.~\cite{Sonia2000} (obtained with the same method). It can be noticed that all $J_{1}^{\textnormal{Co}}$  are similar (differing by  $\sim 6\%$) and larger than in the Ni/Ir$_{n}$/Pt(111) cases  (Fig.~\ref{fig:jDNi}), what indicate a strong ferromagnetic coupling and a different behavior. 
   The presence of Ir spacer layers change the strength of the DM interaction since the largest $D_{1}^{\textnormal{Co}}$ is  for \textit{n} = 0 and  decreases for larger $n$. This can be explained by the smaller induced moment at the adjacent Ir layer compared with the Pt  one (see Table~\ref{tab:mommag_co}).  
 
 The $J_{ij}$ for  Co/Pt(111) from Ref.~\cite{Zimmermann2019} (obtained using the Korringa-Kohn-Rostoker method), agree well to the present results. The small difference found in the case of $J_{1}$
 was also reported in Ref.~\cite{Bornemann2012} ($\sim 7\%$) comparing Co/Pt(111) and Co/Ir(111). We notice that the Co/Ir$_{n}$/Pt(111) $J_{ij}$ and $\vec{D}_{ij}$  behaviors  slightly differ from Ref.~\cite{Vida2016} what  may be related to structural relaxations as well as  different methodologies to calculate $J_{ij}$ and $\vec{D}_{ij}$. 

\begin{figure}[h]
\centering
\includegraphics[width=\linewidth]{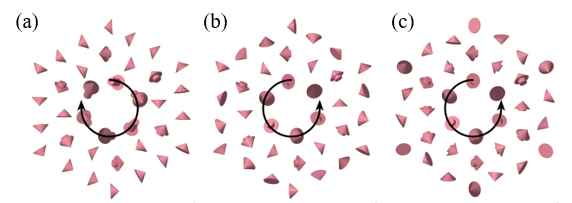}
\caption{\label{fig:dvec_Co} \small{$\vec{D}_{ij}$ 
between the central Co and the Co atom in the site of the arrow in: (a) Co/Pt(111), (b) Co/Ir$_{1}$/Pt(111) and (c) Co/Ir$_{2}$/Pt(111). Arrow sizes are proportional to the vectors strengths.}}
\end{figure}

\begin{figure}[h]
\centering
\includegraphics[width=\linewidth]{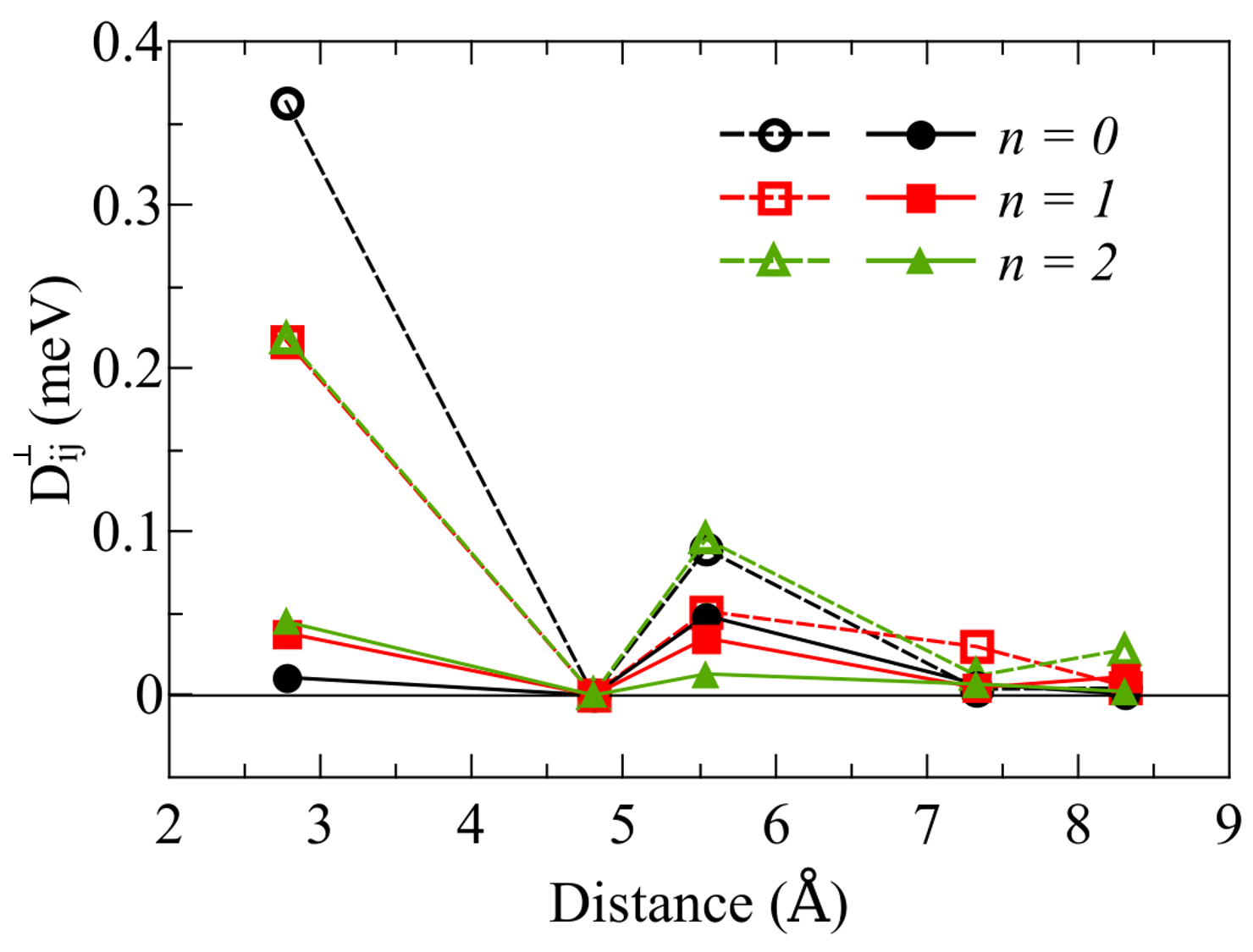}
\caption{\label{fig:dperp} \small{Out-of-plane components of the DM vector ($D^{\perp}_{ij}$) for: Co/Ir$_{n}$/Pt(111) (dashed lines, open symbols) and Ni/Ir$_{n}$/Pt(111) (solid lines, full symbols).}}
\end{figure}

Contrasting to the Ni systems (Fig.~\ref{fig:dvec_Ni}), the DM vector directions in Co/Ir$_{n}$/Pt(111), shown in Fig.~\ref{fig:dvec_Co},  exhibit a change in the rotational sense, with the presence of the Ir spacer layer. Moreover, the $\vec{D}_{ij}^{\textnormal{Co}}$ and  $\vec{D}_{ij}^{\textnormal{Ni}}$ in-plane and out-of-plane components behave differently.
In order to inspect that, in Fig.~\ref{fig:dperp}, the out-of-plane components of the DM vectors ($D^{\perp}_{ij}$) as a function of the distance between two Co or Ni atoms are shown. While $\vec{D}_{ij}^{\textnormal{Co}}$ present a large out-of-plane component, the $\vec{D}_{ij}^{\textnormal{Ni}}$ vectors are mainly in the plane, as presented in \NEW{Sec.}~\ref{subsec:abinitio}.

In order to inspect the 
Heisenberg and DM behaviours, which differ from  Co to Ni based systems, we calculate the $D^{\parallel}_{1}/J_{1}$ ratio, where $D^{\parallel}_{1}$ is the $D_{1}$ component parallel to the plane. The results are shown in Table~\ref{tab:ratio-JD}.
We infer that the existence of skyrmions as ground-state in Ni/Ir$_{n}$/Pt(111), in contrast to a spin-spiral ground-state in Co/Ir$_{n}$/Pt(111), is related to the fact that Co ultrathin layers exhibit smaller $D^{\parallel}_{1}/J_{1}$ ratios, for each \textit{n}, when compared to the Ni-based systems. Atomistic spin-dynamics simulations for Co/Ir$_{n}$/Pt(111) were here performed, resulting, in agreement with Ref.~\cite{Vida2016}, in a spin-spiral  ground-state. 
Nevertheless, since the chirality is introduced when the DM term in the spin Hamiltonian drives a spin rotation to be energetically favorable, we calculated the energy contribution from the Heisenberg and DM interactions in Ni/Ir$_{n}$/Pt(111) ultrathin layers. Here, we define the DM and Heisenberg total energies with respect to a typical site "$0$" ($D^{\textnormal{total}}_{0}$ and $J^{\textnormal{total}}_{0}$, respectively) by

\begin{equation}\label{eq:d_total}
    D^{\textnormal{total}}_{0} = \sum_{j}\vec{D}_{0j}\cdot(\hat{e}_{0}\times\hat{e}_{j})
\end{equation}

\begin{equation}\label{eq:j_total}
    J^{\textnormal{total}}_{0} =  \sum_{j}J_{0j}(\hat{e}_{0}\cdot\hat{e}_{j})
\end{equation},

\noindent where the direction of the magnetic moments ($\hat{e}_{0}$ and $\hat{e}_{j}$) are obtained from spin dynamics simulations. The number of neighbors considered here was 36 (or 5 non-equivalent neighboring shells). The ratio $D^{\textnormal{total}}_{0}/J^{\textnormal{total}}_{0}$ for Ni/Ir$_{n}$/Pt(111) is also presented in Table~\ref{tab:ratio-JD}. It has been  shown in the literature that a larger  $D/J$ ratio favours a faster rotation of the spins, and drives a reduction of the skyrmion size \cite{Fert2013, Sampaio2013, Simon2014}. Our results confirm that, indeed, larger $D^{\textnormal{total}}_{0}/J^{\textnormal{total}}_{0}$ ratios are related to more stable and smaller skyrmions, since the value in Ni/Pt(111) decreases (up to $90\%$) with increasing  Ir buffer thickness.

\begin{table}[h]
	\centering
	\caption{\small{DM/Exchange ratios $D^{\parallel}_{1}/J_{1}$ (see text) and $D^{\textnormal{total}}_{0}/J^{\textnormal{total}}_{0}$ (Eq.~\ref{eq:d_total} and Eq.~\ref{eq:j_total}), for the Co and Ni systems studied here. All ratios are displayed in units $\times10^{-3}$.}} 
	\begin{ruledtabular}
	\begin{tabular}{c|c|cc}
	 & \textbf{Co/Ir$_{n}$/Pt(111)} & \multicolumn{2}{c}{\textbf{Ni/Ir$_{n}$/Pt(111)}}  \\ \hline
	 \textit{n} & $D^{\parallel}_{1}/J_{1}$ & $D^{\parallel}_{1}/J_{1}$ & $D^{\textnormal{total}}_{0}/J^{\textnormal{total}}_{0}$ \\ \hline
 	 0 & 16 & 50 & 4.2  \\ 
    1 & 9 & 130 & 2.9  \\
    2 & 9 & 115 & 0.4  
	\end{tabular}
	\label{tab:ratio-JD}
\end{ruledtabular}
\end{table}


\newpage
\bibliographystyle{apsrev4-2}
\bibliography{bibliography.bib}

\end{document}